\documentclass[sigconf]{acmart}

\usepackage{booktabs}

\graphicspath{{./figures/}}

\renewcommand\footnotetextcopyrightpermission[1]{} 
\newcommand{\boldtitle}[1]{\vspace{5px}\noindent\textbf{#1}}

\newcommand{\eg}{\textit{e.g.}}
\newcommand{\ie}{\textit{i.e.}}
\newcommand{\sj}{\textit{Simjacker }}
\newcommand{\mt}{\textit{MESSAGETAP }}

\usepackage[inline,shortlabels]{enumitem}
\usepackage{tabto}

\makeatletter
\newcommand*{\rom}[1]{\expandafter\@slowromancap\romannumeral #1@}
\makeatother

\usepackage{url}

\usepackage{breakurl}

\usepackage{cleveref}

\usepackage{subcaption}

\AtBeginDocument{%
  \providecommand\BibTeX{{%
    \normalfont B\kern-0.5em{\scshape i\kern-0.25em b}\kern-0.8em\TeX}}}

\setcopyright{acmcopyright}
\copyrightyear{2018}
\acmYear{2018}
\acmDOI{10.1145/1122445.1122456}

\begin{document}

\title{Threat modeling framework for mobile communication systems}


\author{Siddharth Prakash Rao}
\affiliation{%
 \institution{Aalto University}
 \city{Espoo}
 \country{Finland}}
 \email{siddharth.rao@aalto.fi}

\author{Silke Holtmanns}
\affiliation{%
  \institution{Nokia-Bell Labs}
  \city{Espoo}
  \country{Finland}}
  \email{silke.holtmanns@nokia-bell-labs.com}

\author{Tuomas Aura}
\affiliation{%
  \institution{Aalto University}
  \city{Espoo}
  \country{Finland}}
\email{tuomas.aura@aalto.fi}


\renewcommand{\shortauthors}{Rao et al.}

\begin{abstract}
Due to the complex nature of mobile communication systems,  most of the security efforts in its domain are isolated and scattered across underlying technologies.  This has resulted in an obscure view of the overall security. In this work, we attempt to fix this problem by proposing a domain-specific threat modeling framework. By gleaning from a diverse and large body of security literature, we systematically organize the attacks on mobile communications into various tactics and techniques. Our framework is designed to model adversarial behavior in terms of its attack phases and to be used as a common taxonomy matrix. We also provide concrete examples of using the framework for modeling the attacks individually and comparing them with similar ones. 
\end{abstract}

\ccsdesc[300]{Security and privacy~Security requirements}
\ccsdesc[300]{Security and privacy~Mobile and wireless security}
\ccsdesc[300]{Networks~Mobile networks}

\keywords{Threat Modeling, Security framework, Mobile communication}


\maketitle


\section{Introduction}
\label{sec:intro}

Reliance on mobile phones is continuously increasing for purposes other than just making voice calls. More than half of the world population now has a mobile subscription, and a vast majority of it comprises mobile Internet users~\cite{GSMAMobileEconomy19}. While the newer generation of mobile technologies (5G) is slowly paving its way, mobile coverage that relies on previous generations (2G, 3G, and 4G) has increased significantly. It now leaves only 10\% of the world population outside the mobile networks coverage areas~\cite{InternetConnectivity19}. Furthermore, mobile Internet is now affordable to the population from low and middle-income countries~\cite{a4ai19}, and it acts as the first and only means of Internet access. In short, we are moving towards fast, reliable, and robust mobile connections. 
However, what is going at a snail pace is the security of mobile communications. This is because establishing reliable and faster connections has always been the goal of mobile communications rather than achieving secure communication. Security evolved in a closed environment where protocol standardization efforts often require an industry affiliation and software components are proprietary. Up until recent years, network security was achieved by restricting access only to a closed network of trusted partners rather than using strong cryptographic building blocks. Such practices have resulted in a lack of open source modules, tools to conduct audits, datasets about vulnerabilities, and, most importantly, lack of public knowledge. On the contrary, due to the gradual replacement of telephony protocols with IP-based protocols, which evolved in a more open setup, security tools available in the public domain for the latter are now used by hackers to exploit the mobile networks~\cite{GalliumHack}. 

Despite the odds, mobile communication systems have undergone a fair amount of scrutiny from the security research community. For example, internal components of a mobile device
~\cite{golde2013let,guri2015gsmem,samsungbbbkdoor}, 
radio communication between the phone and the cell towers\cite{borgaonkar2011security, shaik2015practical}, 
and mobile core network protocols~\cite{engel2008locating,holtmanns2016user,mashukov2017diameter} have been tested for security-critical issues. All these isolated research efforts are scattered across underlying protocols and technologies, and, it has resulted in an obscure and complex view of mobile communication security. 
While some works have systematized and connected the isolated knowledge, \eg, telephony frauds~\cite{sahin2017sok}, mobile device privacy~\cite{spensky2016sok}, and authentication schemes~\cite{ferrag2018security}, they still look at one specific problem or subsystem. Only a handful of research has taken holistic security into consideration~\cite{rupprecht2018security}. We extend this specific line of research of looking at mobile communication threats as a whole rather than in separate parts. 

Despite its age and plethora of offensive security literature available, mobile communications do not have a domain-specific threat modeling framework. Threat landscape studies and best practice guidelines by standardization and other governance bodies~\cite{GSMAThreatLandscape20, ENISAThreatLandscape19} may use generic threat models, we argue that it is not sufficient. Also, the attacks are communicated mainly in the form of message sequence charts; while they are useful to communicate about a single attack in detail, they do not provide much insight on capturing adversarial behavior that is usually multi-parted. In this work, we fix these shortcomings by proposing a domain-specific threat modeling framework that is built on the existing attacks literature and can be used in parallel with the ongoing development of 5G technology. We believe that our framework initiates a conversation towards unifying prior knowledge and future efforts to secure mobile communication networks.

\boldtitle{Contributions}--- Our contributions from this work are as follows. We provide a comprehensive overview of different subsystems of mobile communications and identify various potential threat actors. Based on a systematic methodology, we present the ``Bhadra framework'' that models the threats to mobile communication systems. In this framework, we categorize publicly known attacks into 8 tactical categories and 47 techniques in total. We organize them further in terms of various phases of the attack life cycle, namely, mounting, execution and results. We also show in detail, with concrete use cases, how to use our framework to capture adversarial behavior by modeling the attacks individually and to compare with similar ones. We also discuss the future directions in which our work can flourish with the help of the mobile communications security community. Our framework is agnostic to underlying technologies and aligned with MITRE ATT\&CK, a popular framework for modeling enterprise IP systems~\cite{strom2018mitre}. Given that 5G networks involve drastic transitions from mobile-specific technologies to IP-specific ones, we believe that the alignment of frameworks makes it easier in terms of future efforts.

\boldtitle{Structure}--- In Section~\ref{sec:motivation}, we describe our motivation for this work. We then give a high-level overview of mobile network topology, including various subsystems and potential adversaries in Section~\ref{sec:background}. In Section~\ref{sec:framework}, we introduce the Bhadra framework, including the methodology that we followed to categorize adversarial tactics and techniques.
We discuss them further in detail in terms of attack mounting, execution and results in Sections~\ref{sec:mounting}, \ref{sec:execution} and ~\ref{sec:resultphase}, respectively. We provide concrete use cases of our models in Section~\ref{sec:usecases}. We discuss the future directions and limitations of our work in Section~\ref{sec:discussion}. Finally, we conclude the paper with closing remarks in Section~\ref{sec:conclusion}.


\section{Motivation}
\label{sec:motivation}

Threat modeling is an essential process in system designs to integrate security and to identify critical aspects of the system that needs to be protected. It is an iterative process that involves defining security requirements as well as identifying and mitigating threats to reduce potential security risks systematically. There are various generic threat modeling frameworks~\cite{shostack2014threat} (\eg, STRIDE, DREAD and PASTA), each with its own advantages and disadvantages~\cite{selin2019evaluation,bodeau2018cyber}. When applying such generic models to a specific domain, it requires careful adaptation or combination with other threat models. For instance, STRIDE is intended for analyzing software vulnerabilities, with access to the source code; however, it is often used for modeling threats, with customization, to distributed systems. Adapting other generic frameworks becomes unmanageable as a system grows mature and complicated. It requires its own threat modeling framework with domain-specific taxonomy and threats. We can find examples of such dedicated threat models, for example, in the domain of storage systems~\cite{hasan2005toward} or industrial control systems~\cite{schlegel2015structured}.

Despite its age and the growing number of threats, there are very few threat modeling frameworks explicitly dedicated to mobile communication systems. The systematization of knowledge (SoK) genre of academic literature about mobile communication as a whole~\cite{rupprecht2018security} or as subsystems ~\cite{sahin2017sok, spensky2016sok, ferrag2018security, rupprecht2018security} points us to the growing needs of the community for a dedicated threat modeling framework. However, to our best knowledge, thesis by Kotapati~\cite{kotapati2008assessing}, which dates back to 2008, is one of the very few works that has attempted to define a threat model for GSM networks. Although the GSM network is still in use today, it co-exists with higher generations of mobile communications and with various components and features that did not exist in 2008. Nonetheless, our primary motivation is to extend domain-specific threat modeling for mobile communication systems, that is agnostic to the underlying technologies, and can be used to model the threats from all possible attack surfaces that are part of the newer generations.

Our other motivation comes from the requirements of a large mobile network provider company. The company comprises a wide range of employees: engineers who build end-to-end mobile communication systems, technical sales and marketing executives who sell the products and services to mobile operators, researchers who contribute to both existing and future solutions, standardization experts who exchange knowledge by participating in standards committees, and other technical support personals who assist customers with fixing technical issues. They share equal responsibilities for securing mobile communication systems altogether. Despite having in-depth technical knowledge required for their respective roles, the main problem they currently face is the lack of a common taxonomy and metrics to capture a high-level overview of the state of security of the entire system. The company uses the MITRE ATT\&CK threat modeling framework~\cite{strom2017finding,strom2018mitre} on the enterprise management side, and they believe a dedicated framework for mobile communication systems that can co-exist with ATT\&CK will be useful.

Due to the complex nature of the mobile communication systems, we also believe that the lack of a common taxonomy to communicate security-related issues is a generic problem amongst most of the companies working in this sector. These companies heavily rely on the resources produced by the 3\textsuperscript{rd} Generation Partnership Project (3GPP) in the form of normative technical specifications (TS) and informative technical reports (TR). For example, series 33~\cite{3GPP33Series} and 35~\cite{3GPP35Series} provide in-depth knowledge of individual subsystems, in terms of how the underlying technology can be built securely. Other regulatory bodies such as the GSM Association (GSMA) ~\cite{GSMASecurity} and the National Institute of Standards and Technology (NIST)~\cite{cichonski2016guide} also produce security studies and guidelines that complements 3GPP's efforts. They also produce resources that discuss the summary of attacks on mobile communication systems~\cite{GSMAThreatLandscape20,franklin2016assessing, ENISAThreatLandscape19}. Although all these resources provide a common taxonomy and attack categorization of some sort to the companies in the mobile communication sector, they unfortunately cannot replace traditional threat models. 

Furthermore, both the technical specifications and attack literature use message sequence charts (also known as sequence diagrams) as a standard form of communicating the working mechanisms and related vulnerabilities. Given the protocol-heavy nature of the mobile communication systems, such message sequence charts provide a clear understanding of how protocol messages between different network entities work or can be exploited. However, they fail to capture the multi-parted adversarial behavior of an attack or to provide any further insights than can be derived while studying multiple attacks of similar kinds. We believe that this can also be fixed with an appropriate format of the threat model.

To this end, our goal is a design a domain-specific framework that provides a common taxonomy and categorization of attacks and retains the usefulness of message sequence charts but in a more insightful manner. Our goal is also to ensure that the threat model is simple and easy to be used by different technical roles of a company.


\section{Background}
\label{sec:background}
\begin{figure*}
  \includegraphics[width=\textwidth]{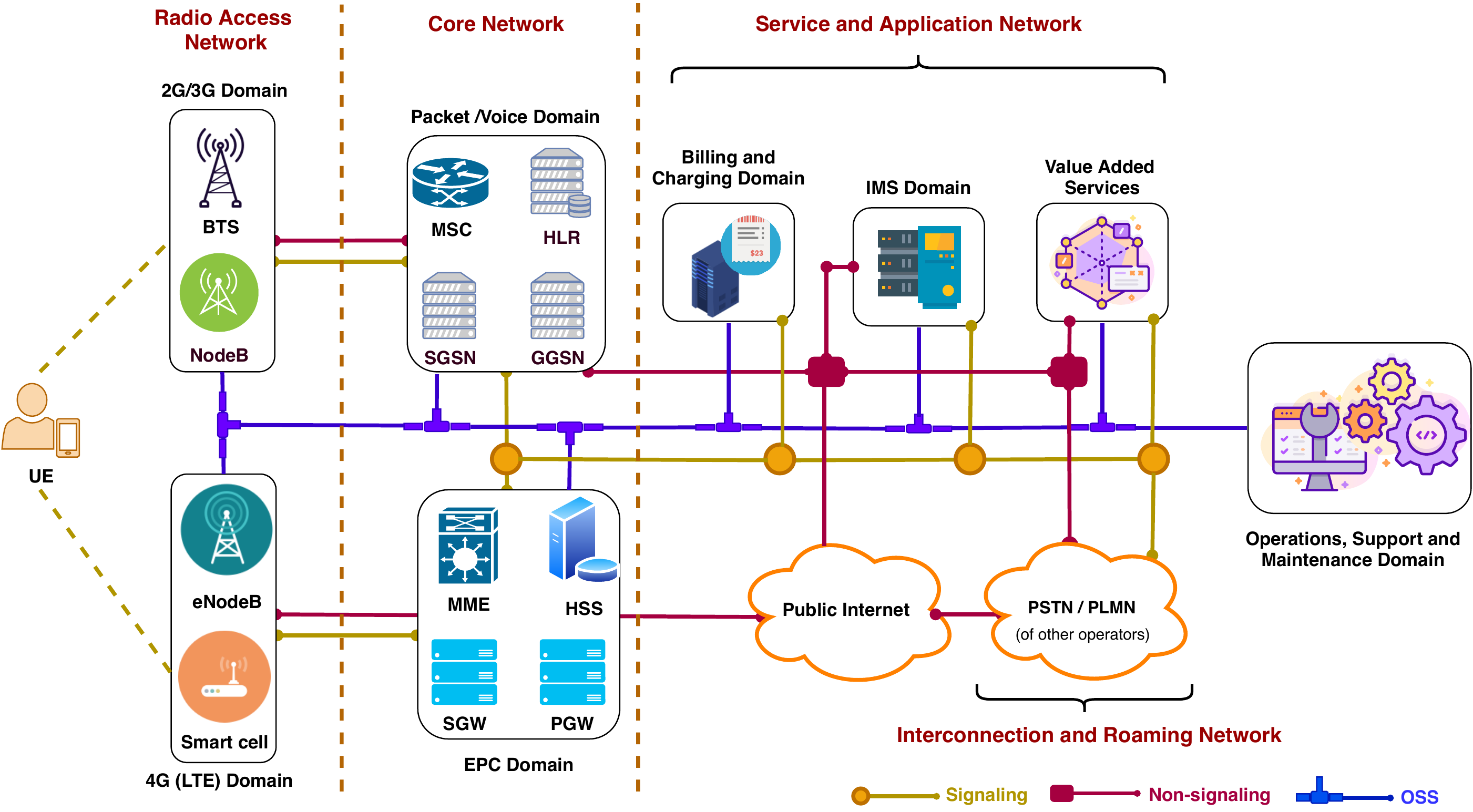}
  \caption{Overview of mobile networks topology}
  \label{fig:topology}
\end{figure*}


Significant changes in the nature of technology and architecture in mobile communication networks are collectively referred by their standardized ``generation (G)''. Each generation is expected to overcome the limitations of their previous generations with improved network capabilities (\eg, speed, frequency, data capacity, security, and latency) and with new techniques and features. The first generation (1G) used analog network signals, which could support only voice calls. Digital signals are used from the second generation (2G) or Global System for Mobile Communications (GSM) onward, which introduced Short Message Service (SMS) and Multimedia Messaging Service (MMS), the latter although being laggy and often without success. Intermediate generations such as the General Packet Radio Service (GPRS or 2.5G) and Enhanced Data rates for GSM (EDGE or 2.75G) improvised network capabilities to enable stable internet connections, especially when the user is on the move. Both 1G and 2G use circuit switching connection-oriented networks, with a dedicated route between the source and destination until the entire message is transferred through it, for both voice and data transfers. In the third generation (3G) or Universal Mobile Telecommunications System (UMTS), connection-less packet switching networks improved speed and reliability for transmission of larger amounts of data. With that, mobile users could make video calls, stream media, and play online games from their 3G enabled mobile phones. The fourth-generation (4G) or Long Term Evolution (LTE) increased the bandwidth and lowered latency of mobile Internet connection, which contributed largely to the growth of mobile broadband with a high quality of streaming (\eg, less buffering than 3G connections). Both voice (through Voice over LTE) and data are transmitted through IP-based packet switching networks in 4G. We recommend that the readers refer to the work of Rost et al. \cite{rost2016mobile} for more details about the evolution of mobile network architecture.

\subsection{Mobile networks topology}
We now present a high-level overview of modern mobile network topology (refer to Figure~\ref{fig:topology}) and describe functionalities of the associated components. This serves as the technical background required for the rest of the paper. Please note that we mainly focus on the generations of mobile communication that currently co-exists with each other at the time of writing this paper, \ie, 2G, 3G, and 4G only.

\subsubsection{\textbf{User Equipment (UE)}}
\label{subsec:ue}
User Equipment (or mobile stations in GSM networks) --- the mobile device used by mobile subscribers for using mobile services --- is equipped with integrated circuit Subscriber Identification Module (SIM) physical smart cards. Modern smart cards are called the Universal Integrated Circuit Cards (UICC). We use the term UE and SIM card unanimously to refer to the mobile device and smart cards, respectively, of all mobile generations. The UE contains all the hardware and software, and the SIM card includes a mobile subscription profile and the cryptographic keys needed for communication with mobile networks. Each SIM card is identified by its International Mobile Subscriber Identity (IMSI). Furthermore, each SIM card slot is associated with a unique identifier called International Mobile Equipment Identity (IMEI), which is used for registering the device to the network. So, mobile SIMs with dual SIM feature will have two IMSI and IMEI pairs. 

Both IMSI and IMEI are transmitted over the air to establish radio channels between the UE and the mobile network, and they remain a secret between the user and the mobile network. On the other hand, the Mobile Station International Subscriber Directory Number (MSISDN, i.e., the phone number) is the only public identifier that users can share with their friends and family. The IMSI, MSISDN, and IMEI are unique and permanent identifiers in mobile communication, and they will change only when the user changes its mobile subscription or equipment. Mobile networks require to frequently check the subscription of the user to allow it to use the network continuously. To avoid the repetitive use of IMSI, provisional, and short-lived identifiers such as Temporary Mobile Subscriber Identity (TMSI) or Globally Unique Temporary ID (GUTI) are used as alternatives. 

\subsubsection{\textbf{Radio Access Network (RAN)}}
\label{subsubsec:RAN}
The radio access network is the first point of network access, which wirelessly connects the UE over the air to the core network that provides telephony services. This segment comprises cell towers of Base Transceiver Station (BTS) and NodeB from 2G and 3G domains, respectively. Furthermore, the evolved NodeB (eNodeB) and smart cells of the 4G (LTE) domain provide air connectivity with additional features such as support for voice over WiFi. Since the radio access network is responsible for keeping the connection to the rest of the network intact without disconnecting, including when a user is on the move, it is designed to transition between different generations through handover mechanisms.

\subsubsection{\textbf{Core Network (CN)}}
\label{subsubsec:CN}
The core network is responsible for managing mobility of the users by interacting with the RAN, for initiating connections with other network operators and delivering telephony services (such as voice calls, SMS, and Internet data connections) requested by the users. As represented in figure~\ref{fig:topology}, we consider only the critical nodes from the packet/voice domain from 2G/3G and the Evolved Packet Core (EPC) domain as part of CN. 

\textit{Home Subscriber Server (HSS)} from 4G/LTE EPC domain is a master database that maintains user information (\eg, IMSI and MSISDN) and their subscription information in one single node.  HSS is responsible for user authentication and access authorization of what services can the users request based on their subscription plans. HSS is also in charge of mobility management and supporting call/data session establishment, for example, by keeping track of the user's whereabouts and relaying it to other nodes. \textit{Home Location Register (HLR)} is the counterpart of HSS from the packet/voice domain. However, HLR requires a separate node called Authentication Center (AuC) for user authentication, whereas it comes as an integral part of HSS.

\textit{Serving Gateway (SGW)} and \textit{Packet Data Network Gateway (PGW)} from the EPC domain are the user plane gateway nodes which route and filter the IP traffic between the UE and the external networks. More specifically, SGW is the point of interconnection between the RAN and CN, whereas PGW is between CN and external roaming/interconnection network, including the public internet or other mobile operators. They both are in charge of supporting accounting and charging of services, managing user mobility, and providing lawful interception for the user plane. \textit{Gateway GPRS support node (GGSN)} and \textit{Serving GPRS support node (SGSN)} perform similar functionalities in the 2G/3G networks.

\textit{Mobility Management Entity (MME)} handles the control plane traffic in 4G networks. More specifically, it handles signaling related to session and mobility management, user authentication (with the help of HSS), and selection of the gateways. MME is also responsible for the lawful interception on the control plane. The \textit{Mobile Switching Center (MSC)} is responsible for similar functionalities in the 2G/3G networks.

\subsubsection{\textbf{Service and application network (SAN)}} 
\label{subsubsec:SAN}
This network comprises of the billing and charging domain, IP multimedia subsystem (IMS), and value-added services (VAS). The billing and charging domain is usually considered as part of the core network in the mobile network literature, as it was mainly used for accounting the voice call usage in a standard manner~\cite{kuhne2011charging}.
While voice call services are still part of the core network, current mobile networks have evolved much beyond that, for example, to bill for the usage through IMS and VAS networks. 

 The IMS domain integrates mobile and fixed voice communications with Internet technologies. One of the key features of IMS is the standardization of Session Initiation Protocol (SIP)~\cite{RFC2543}, which enables third-parties (who are not mobile operators) to provide voice and media services over the IP-to-telephony networks. Similarly, mobile operators partner with various third-party vendors to offer a wide range of value-added services, such as missed call and voice box services, mobile commerce and advertisements, gaming, and on-demand streaming.  Mobile commerce is another such partnership where the mobile users can buy commodities from registered third-party vendors, and the value of the commodity is charged towards the user's mobile subscription. Such partnerships increase revenue to the operators and provide a large user base to the vendors in exchange for the operator's service fee.

\subsubsection{\textbf{Operations, support \& maintenance network (OSMN)}}
\label{subsubsec:OSMN}
Most operators support multiple generations (from the modern 4G to legacy 2G) of mobile network infrastructure. Given the heterogeneous and complex nature of such networks, the mobile operators collaborate with external vendors to manage, configure, and monitor their networks. Such vendors are known as the Operations Support System (OSS) in the literature; however, we refer to them as ``Operations, support and maintenance network'' (OSMN) to indicate their broad range of functionalities. OSMN requires a connection to every node from RAN, CN, and SAN for managing and troubleshooting purposes. 

\subsubsection{\textbf{Interconnection and roaming network (IRN)}}
\label{subsubsec:IRN}
So far, we have described the different kinds of network subsystems that belong to a single mobile network operator. These network subsystems are collectively referred to as the Public Switched Telephone Network (PSTN). Each operator owning such PSTN networks, along with their fixed Public land mobile networks (PLMN), communicates with one another to provide seamless mobile communication between their respective users. Also, operators communicate with each other during ``roaming'' where a user goes out of its home operator's geographical coverage area and visits of another operator. In such cases, the user will still be able to use its mobile services through the network of the visited operator.  We use Interconnection and Roaming Network (IRN) as a generic term to refer to such operator-to-operator communication. 

Similarly, each mobile network also has to provide internet connections to its users, for example, for mobile browsing, SIP calls, and other IP-based communication. In such contexts, the core network connects service and application networks to the public Internet. In such operator-to-internet connections, what comes towards the operator's network is outside the control of the mobile operators.

\subsection{Communication between the networks}
We now describe how the aforementioned networked subsystems communicate with each other or within themselves. To limit the complexity, we only provide a high-level overview with relevant communication protocols used by each subsystem.

\subsubsection{\textbf{UE with RAN}}
\label{subsubsec:UEwithRan}
User equipment connects with base stations over radio channels where both the UE and base stations exchange generic broadcast and paging messages to inform each other about their whereabouts. A dedicated radio channel is established when a call, SMS, or browsing sessions are initiated. A challenge-response Authentication and Key Agreement (AKA) protocol is also run over the radio channel to derive session keys between the UE and the network. Only the UE authenticates itself to the network in GSM networks, whereas in 3G and 4G, AKA protocol is extended to provide mutual authentication.

\subsubsection{\textbf{Core network with other networks}}
\label{subsubsec:CNwithOthers}
Most of the interactions that the CN does with other nodes involve signaling, a term that refers to the use of signals for controlling communications, for mobility management and call establishment.
The 2G and 3G networks use Signalling System 7 (SS7) or SIGTRAN (the adaptation of SS7 over IP) as signaling protocols. The SS7 protocol stack was developed in the days of fixed landlines to exchange information among different nodes of the same operator or between operators and eventually adapted to mobile communication. While SS7 and SIGTRAN are replaced by Diameter protocol in  4G networks, SS7 is still the extensively used signaling protocol today due to the dominance of 2G networks around the globe. Inter-generation signaling communication (2G/3G to 4G or vice versa) is facilitated by Inter Working functions (IWF) on the edge nodes. 

GPRS Tunnelling Protocol (GTP) is another signaling protocol that is common across all generations of mobile networks for signaling related to maintaining a data connection (i.e., for internet access) while on the move and for carrying the data.

\subsubsection{\textbf{OSMN with other networks}}
\label{subsubsec:OSMNwithOthers}
OSMN relies mainly on the Common Management Information Protocol (CMIP) and Simple Network Management Protocol (SNMP) to provide remote network configuration and management of the operator's nodes. Other popular protocols such as the Secure Shell (SSH),  File Transfer Protocol (FTP), Virtual Private Networks (VPN), Representational State Transfer (REST) and Simple Object Access Protocol (SOAP) are also commonly used to provide the OSMN with command-line or user interfaces for administering the nodes. 

\subsubsection{\textbf{Transport networks}}
\label{subsubsec:TN}
IP-based network transport is used extensively in mobile communication systems, for example, while establishing a data service from the UE to an IP endpoint (such as a web server), or while using signaling protocols over an IP relay network. Stream Control Transmission Protocol (SCTP) is used as a transport layer protocol between the nodes within an operator.  In the case of roaming or communication with other operators, the guidelines for inter-service provider IP backbone (as per IR.34~\cite{IR34}) recommends using Internet Protocol Security (IPSec) or other VPN connectivity. However, instead of VPNs, they use NAT middleboxes for separating private networks of mobile operators from the public Internet. Although both the user (data) and control (signaling) traffic is separated from each other, misconfigurations or security loopholes in the private networks may allow non-operators to tamper with such traffic.



\subsection{Potential adversaries}
\label{subsec:adversaries}
Based on capabilities and type of access to the different subsystems of mobile communications, we consider the following potential adversaries.

\subsubsection{\textbf{Radio link attackers}}
\label{subsubsec:RadioAttacker}
Inexpensive hardware~\cite{EttusSDR} and open source software modules ~\cite{gomez2016srslte} have proliferated security research of radio communication outside the mobile operations industry. However, they have also given a chance for evil actors to build tools and products that undermine the security of radio channels, which otherwise would have been not possible. With such tools, radio link attackers can exploit the fact that mobile phones have no way of authenticating legitimate base stations during their initial connection establishment phase~\cite{van2016effectiveness,jover2016lte,park2019anatomy}.

\subsubsection{\textbf{Evil mobile operators}}
\label{subsubsec:MobOperator}
As per the current standards, the phone encrypts the communication between itself and the radio access network using the keys stored on the SIM card.
Encryption beyond radio networks can exist, but only in the form of data transmission inside encrypted tunnels. Mobile operators need to route the communication to other operators, and both operators participating in the communication will have access to unencrypted data inside the tunnel if it exists. Such access is an accepted norm since the inception of global mobile communication systems because the interconnection protocols (\eg, SS7) were built for mutually trusting government-owned mobile operators. Now that things have changed, most of the mobile operators are private entities, and the government itself can be a potential threat. Having access to unencrypted data and control over routing gives mobile operators the ability to impersonate other trusted operators easily.

\subsubsection{\textbf{Human Insiders}}
\label{subsubsec:Insider}
Humans are always considered as one of the weakest links in system security because they are prone to make mistakes. Employees of mobile operators or OSM networks can go rogue to exploit their privileges to leak sensitive data~\cite{NotSoSecurus}, for the sake of their ideologies (\eg, whistle-blowing), or sell them for financial gain~\cite{ss7ForSale}. Carelessness and lack of security education can also make human insiders gullible, \eg, to social engineering, misconfiguring nodes, or to disregard operational security. In any case, human insiders, with their direct access to critical infrastructure, are potential adversaries.

\subsubsection{\textbf{Hardware and SIM manufacturers}}
\label{subsubsec:HWandSimMakers}
Manufacturer of network nodes and phone hardware with evil intents can be potential attackers as they can induce threats in the hardware supply chains. Bugs at the hardware-level are challenging to trace and coming to light only recently ~\cite{robertson2018big}. SIM card manufacturers also pose similar threats by distributing SIMs with buggy features or backdoors. On the other hand, SIM cards also contain private keys used for encrypting the over-the-air radio communication. Careless management of the key generation infrastructure or cooperation with other actors (\eg, oppressive governments) pose a serious threat to mobile communication~\cite{scahill2015great}.

\subsubsection{\textbf{Software and OS vendors}}
\label{subsubsec:SWandOSvendors}
Mobile networks involve a large number of open-source or proprietary software components to enable the regular functioning of the communication. Similar to hardware, the software supply chain is also prone to contain intentional or accidental vulnerabilities. Due to its complex and closed nature, breaking into the core network requires in-depth knowledge and skills, whereas, with publicly available forensic tools, attackers can exploit common vulnerabilities (\eg, SQL injection~\cite{tung2014hackers}) in the software stack of routers and other network devices~\cite{hau2015synful,checkoway2016systematic}. Since software and OS vendors may become a medium of software supply chain infiltration, we consider them as potential adversaries.

\subsubsection{\textbf{Law enforcement and oppressive governments}}
\label{subsubsec:lea}
Legal entities such as the law enforcement agencies have separate interface standards for lawful access to mobile communication systems~\cite{li2018comprehensive,LIdocsETSI}, and every mobile operator has to support it based on the laws of a nation. Nonetheless, such entities have exploited mobile communication data outside the lawful interfaces, \eg, in mass surveillance programs~\cite{gellman2013nsa} and malware campaigns ~\cite{ReginGSM}. Considering the power and interest of nation-state actors in obtaining access to the internal networks of mobile operators, including infiltrating into hardware or software supply chains, we treat them as potential attackers to mobile systems.

\subsubsection{\textbf{Evil mobile users}}
\label{subsubsec:evilusers}
Most mobile phones contain an application processor, for running the mobile OS and general user applications, and a baseband processor for the radio software stack involving communication over cellular networks. The former is usually open-sourced and freely available for users to run mobile apps of their choice. The baseband processor, however, is proprietary, and accessing them requires reverse engineering. Nonetheless, mobile users have access to both of these. On the one hand, a skillful user can build apps, \eg, to modify traffic (mostly web) transmitting out of the UE to gain free data services. On the other hand, he can also tamper with the baseband processors, \eg, to spoof its identity to the mobile network. Given the UE-side tampering capabilities, we treat skillful mobile users with evil intent as potential adversaries.


\section{Threat Modeling Framework}
\label{sec:framework}
In this section, we introduce the ``Bhadra'' threat modeling framework. More specifically, we describe the methodology that we followed and our design choices. We also define the building blocks (\textit{i.e}., the \emph{tactics} and \emph{techniques}) of our model. This section serves as a preliminary to the rest of the paper.

\subsection{Methodology}
\label{sec:methodology}
\begin{figure}[ht]
  \includegraphics[width=\columnwidth]{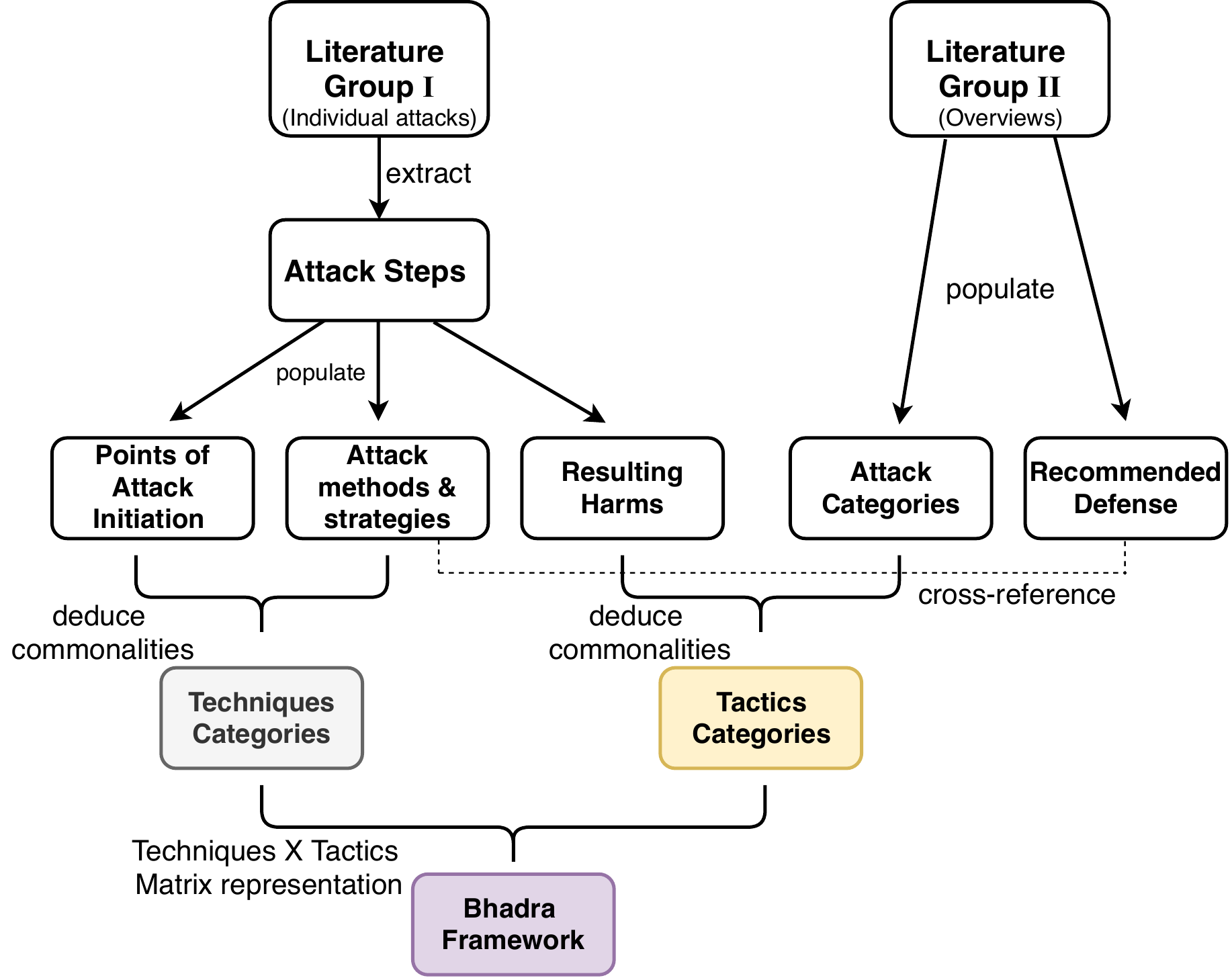}
  \caption{Overview of methodology}
  \label{fig:methodology}
\end{figure}

The methodology that we followed to develop our threat modeling framework is represented in Figure~\ref{fig:methodology}. We used the existing attacks and defenses literature from the following two groups as references to build our framework. 

\begin{itemize}
    \item \emph{Group \rom{1}:} This group comprises of peer-reviewed academic publications and presentations at information security conferences. It offers a rich resource of individual attacks in-depth about the attack flow and root cause.
    \item \emph{Group \rom{2}:} The literature in this group is curated by standardization bodies from the mobile communication sector (\textit{e.g.}, 3GPP, GSMA, ETSI) and government agencies (\textit{e.g.}, ENISA and NIST). It contains a generic summary of subsets of attacks as well as best practice guidelines and recommendations for building defensive strategies. 
\end{itemize}

Firstly, we extracted the \emph{``attack steps''} from the literature group~\rom{1}, which is usually available in the form of message sequence charts and followed by brief descriptions. Then, we populated the \emph{``points of attack initiation''}, \emph{``methods and strategies''} of an attacker, and the \emph{``harms''} caused from each attack. Similarly, we populated the generic \emph{``categories of attacks''} and \emph{``recommended defense''} strategies from literature group~\rom{2}.

Secondly, by finding commonalities in the points of attack initiation and attack methods, we grouped them into abstractions of different \emph{``categories of techniques''}. We cross-referenced the defense strategies (populated from group~\rom{2}) with the categories of techniques to validate whether an attack execution bypasses any of the recommended defenses. This cross-referencing led to deduce a techniques category called \emph{``Defense evasion"}. Similarly, by deducing commonalities between resulting harms (from group~\rom{1}) and attack categories (from group~\rom{2}), we group them into different \emph{``tactical categories''}. While defining the tactical categories, we aligned them with the MITRE ATT\&CK framework, which indeed was our initial requirement. We could retain names of tactics from ATT\&CK with only a few modifications to customize it into the context of mobile communications. Finally, we represent the techniques as rows and tactics as column headers to form the \emph{Bhadra framework}.

\boldtitle{About our design choice. } 
The ATT\&CK framework focuses on documenting common tactics, techniques, and procedures of malware and advanced persistent threats to build a knowledge base of adversary's offensive behaviors through attack life cycles against particular platforms (\eg, Windows). It is based on real-world observations gathered, \eg, through malware samples, penetration testing, and threat intelligence reports. It focuses on how adversaries interact with the system during an attack rather than the tools or malware. Such behavioral modeling of attackers from the previously known attacks helps to recognize the responsible adversary groups~\cite{AttckGroups} and defend against them during the early phases of an attack. 

Our preference for aligning the Bhadra framework with the ATT\&CK might seem biased to favor the requirement of the mobile service provider company, which uses it for their enterprise management side (refer to Section~\ref{sec:motivation}). However, that is not the case. 
Unlike the attacks considered by the ATT\&CK framework, the attacks on mobile communication systems are not conducted by any publicly known adversary groups. Most attacks are presented by academic researchers and information security professionals. Nonetheless, by studying the message sequence charts of the attacks, we believe that modeling the adversary behavior as if it is from a specific group is still useful. This is because the attack flows, which can be mapped to adversary's behavior, indicate the limitations on how adversaries can compromise the system and the loosely protected aspects that need more rigorous security. Also, the matrix representation not only helps in building a systematic categorization and common taxonomy based on the known attacks but also to retain the attack flows of a message sequence chart. 

\subsection{Tactics and techniques}
\label{subsec:tacticsandtechniques}
We consider the end-to-end communication over the mobile networks as the primary asset to be protected. In this realm, all entities facilitates the communication, right from the user equipment to all the network nodes (represented in Figure~\ref{fig:topology}), are considered as the assets to be protected. We consider any scenarios where functionalities of the nodes are misused or abused, which could incur financial or accounting discrepancies to a mobile operator as valid threats. We also consider the attacks that affect the mobile users, say harm privacy of their communication, or incurs them a fee for services that they have not used. However, we omit the threats from malicious apps or malware on their devices that steal sensitive information (\textit{e.g.}, banking credentials, or passwords to online services) but do not affect mobile communication as such. The threat scope is abstracted in the form of various tactics and techniques that act as the building blocks of our framework. They are defined as follows.

\boldtitle{Tactics} represents the adversary's tactical goals, \textit{i.e.}, the reason (``why'') for performing a particular action during an attack. As contextual categories of underlying techniques, tactics represent the types (sub-phases) of adversarial actions right from the beginning of an attack until it ends. We organize the tactics in the way they represent the natural attack strategy of an adversary. More specifically, the adversary finds (or knows) a weak point to mount the attack, he then executes the attack, and finally achieves his objectives by gathering the desired results. Most of the attacks have one or more techniques from each of the tactics. However, some attacks may skip some of the intermediate tactics and corresponding techniques.
    
\boldtitle{Techniques} refers to the ``how'' and ``what'' aspect of adversarial strategy and adds more context to adversarial tactics. Techniques could represent individual actions (grouped by their nature) taken by an adversary to achieve its tactical objectives. It could also refer to the information that the adversary learns by performing a specific action. There are different types of actions (or information that can be learned from an action) that an adversary can perform to achieve each of his tactical objectives. This is why we have multiple techniques under each tactical category.

While deducing tactics and techniques for our framework, we referred to both enterprise and mobile categories of ATT\&CK framework and adapted them as per our needs. The techniques in our model are specific to mobile communications, and the tactics are aligned with the ATT\&CK framework with minimal modification as follows. Firstly, we dropped the \emph{execution} and \emph{privilege escalation} categories because we do not have sufficient public information on specific techniques that can fall under that category. Nonetheless, network nodes and routers still run common operating systems (e.g., Linux), and all the techniques under execution and privilege escalation tactical category from ATT\&CK may still be valid here. We group them into a single technique called ``exploit platform- and service-specific vulnerabilities''. Secondly, the \emph{commands and control} tactic contained techniques that adversaries may use to communicate with systems under their control, for example, by mimicking normal, expected traffic. We aptly used the same definition; however, aptly renamed it as \emph{standard protocols} due to its prevalent use in the mobile communication community. Thirdly, we combined \emph{credential access}, \emph{exfiltration} and \emph{collection} into a single tactic called \emph{collection}. Finally, we also combined \emph{impact} from enterprise domain, \emph{network effects} and \emph{remote service effects} and mobile domain to a single category called \emph{impacts}. We also modified its definition to refer to the highest level of the result achieved by the adversary. 

\begin{figure*}
  \includegraphics[width=\textwidth]{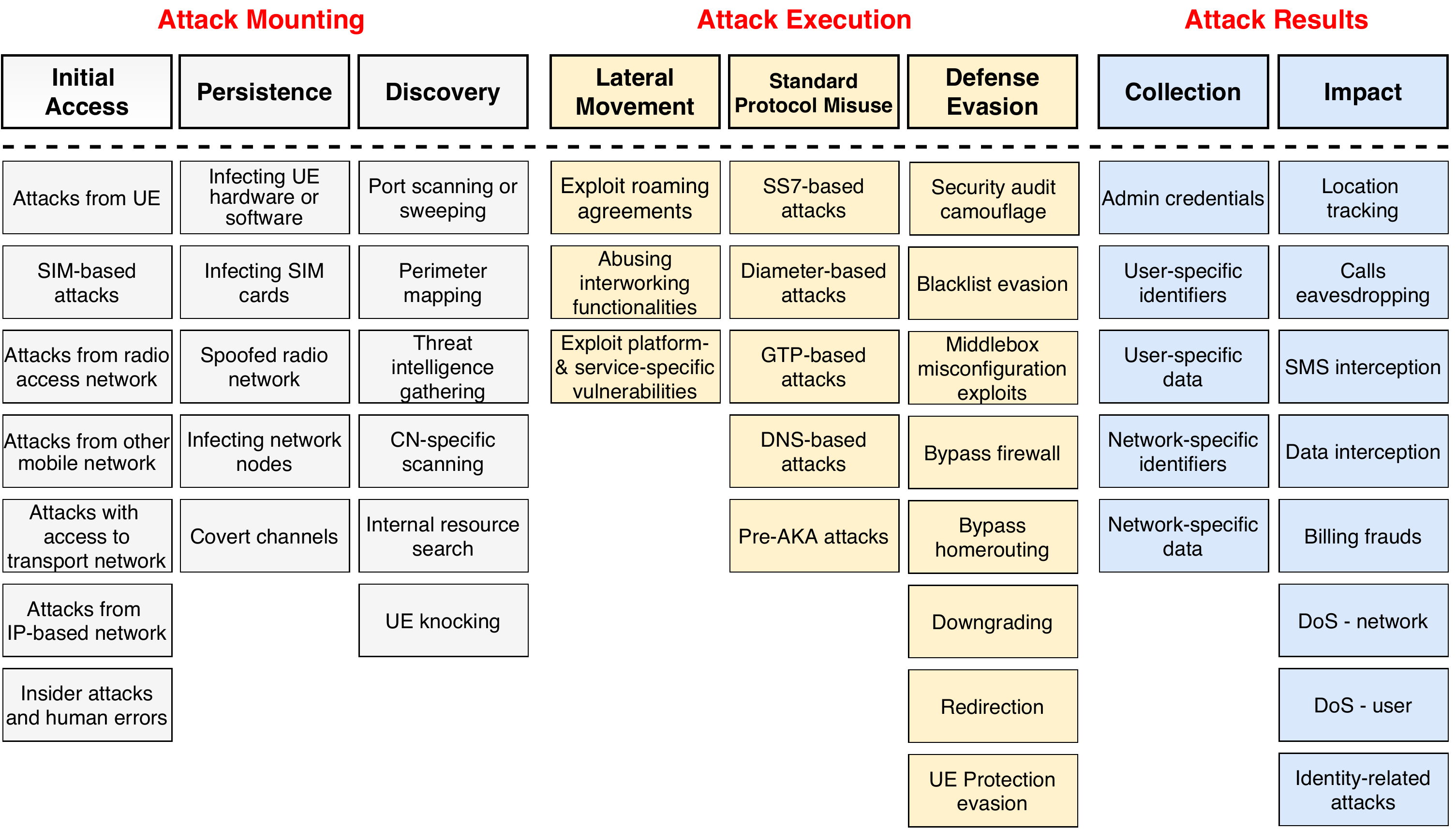}
  \caption{Bhadra threat modeling framework}
  \label{fig:bhadraframework}
\end{figure*}

We describe the tactics and techniques categories in terms of attack mounting, execution and result gathering phases of the attack life cycle in Sections~\ref{sec:mounting}, \ref{sec:execution} and \ref{sec:resultphase} respectively. In each of these sections, we describe tactics and underlying techniques (emphasized in \textbf{``bold''}) in detail with relevant examples and references.  With all tactics and techniques, Figure~\ref{fig:bhadraframework} represents the Bhadra framework.


\section{Attack mounting phase}
\label{sec:mounting}
The first phase of the attack life cycle is the attack mounting where the adversary finds a weak point as its target to mount the attack and ensures that its control persists as long as it is required. The adversary may also gain more information to prepare for the next phases. We recognize three tactics in the attack mounting phase, namely, initial access, persistence, and discovery. 

\subsection{Initial Access}
\label{subsec:IntialAccess}
Initial access represents the group of techniques or attack vectors that the adversaries use as entry points, \eg, by exploiting weaknesses in the system or by luring humans with access to the system. In its entirety, the techniques under the initial access tactics represent various entry points through which adversaries can launch attacks on the mobile communication systems.  
%

``\textbf{Attacks from UE}'' refers to any technique that involves the attacks launched by the software or hardware components of the user equipment to send malicious traffic into the mobile network. We see mobile malware as the malicious apps installed on the UE that often compromise the mobile user's privacy by stealing sensitive information or capturing the user's activities. However, there are other kinds of malware observed in the wild that affects mobile communication. A large number of UEs infected with such malware can form cellular bots and overload the critical nodes (\eg, HLR/HSS) of the operator~\cite{traynor2009cellular}. Massive traffic overload from the user endpoints, especially from IoT devices, is an expected threat in the 5G network, and there are solutions proposed to fix it~\cite{salva20185g}. Besides malware, skillful mobile users can also inject malicious traffic into the mobile network. More specifically, mobile users can create malformed IP traffic emerging from the UE, \eg, to avail free data services~\cite{peng2012mobile}. Such attacks were popular when there was a much stricter cap on mobile data usage. 

The ``\textbf{SIM-based attacks}'' are the techniques that involve any physical smart cards, namely SIM from 2G, USIM from 3G, and UICC from 4G networks. The most well-known examples of SIM-based attacks are the swapping~\cite{lee2020empirical} and cloning~\cite{anwar2016forensic} techniques. SIM swapping attacks work by convincing customer service representatives of a mobile operator, \eg, at a local retail store, to give a replacement SIM for a subscription that is still in use. In legitimate scenarios, SIM replacement is valid to help mobile users who have lost their SIM cards. The representative authenticates the person who claims to be the previous owner of the SIM based on correct answers to questions about personal details or call records. Such human-to-human authentication is not secure by any means. Also, there is a big market to recruit customer service representatives of mobile operators~\cite{SIMSwapRecruit} to conduct SIM swapping at a large scale. Successful acquisition of a SIM gives the attacker access to the IMSI and master encryption key, which can be used by the attacker to impersonate the mobile subscriber to the network. SIM cloning (or rooting) has a similar goal as SIM swapping to acquire the IMSI. However, the attacker relies on physical access to the SIM to extract all contents and to make a replica. SIM cloning can also be conducted remotely in certain specific scenarios~\cite{nohl2013rooting}. A relatively new and lesser-known type of SIM-based attack is exploiting functionalities of the SIM card. We discuss more in detail about one such recent attack called \emph{SIMjacker}~\cite{SIMjacker} in Section~\ref{subsubsec:simjacker}.

The ``\textbf{attacks from radio access network}'' are the techniques where an adversary with radio capabilities impersonates the mobile network to the UE (or vice versa) and becomes a man-in-the-middle. The main requirement for these attacks to work is that the adversary has to be present within the radio range of the victim so that the victim's UE would pick up the adversary's radio signals instead of benign signals from the operator's cell towers (base stations). A popular example in this category is the use of IMSI catchers (also known as stingrays) where an adversary steals IMSIs (or phone numbers in some cases) from the UEs within the adversary's close vicinity~\cite{borgaonkar2011security, shaik2015practical, ShaikIMSICatcher16,park2019anatomy}. Depending on the generation of mobile communication, such adversaries can intercept the communication (\eg, SMS and calls) emerging from the UE, can act as jammers by denying service, and track the presence of a specific UE in a given location~\cite{GottaCatchThemAll}. Since the communication interception is limited only to 2G networks, the adversary in most cases may simply want to downgrade 3G and 4G connections to 2G. Nonetheless, IMSI catching attacks without downgrading are still possible in some cases~\cite{borgaonkar2015lte}. Attacks using femtocells (home nodeB) work similar to IMSI catchers but mostly in 3G radio networks~\cite{borgaonkar2011security,golde2012weaponizing}.

The ``\textbf{attacks from other mobile networks}'' and the ``\textbf{attacks with physical access to transport network}'' techniques can be conducted by evil mobile operators, law enforcement agencies for legal interception and human insiders with access to network nodes. 
The major difference between these two techniques is that the former attack technique is launched from a partner mobile operator via IRN, mainly during roaming scenarios, and the latter can also be launched from the victim's own operator's core network. In either case, the adversaries mostly rely on launching messages of standard cryptographic protocols for location tracking, communication interception, denial of service, billing frauds~\cite{rao2015analysis}. Unlike the attack techniques with the radio access network, the adversary's presence within the radio range of the victim is not required. Instead, the adversary can remotely conduct the attacks at a larger scale from any corner of the world.

The ``\textbf{attacks from IP-based attacks}'' techniques mostly are launched from the service and application network, which allows non-operator entities to infuse malicious traffic into an operator's network. Operators become vulnerable to the entire arsenal of Internet-based attacks~\cite{keromytis2011comprehensive}, including attacks on inter-domain routing~\cite{butler2009survey}. Most commonly, such techniques aim for billing frauds~\cite{zhang2007billing} and denial of service~\cite{sisalem2006denial,ehlert2010survey} using SIP protocols. The GPRS traffic emerging from the interconnection and roaming network can also give initial access to the adversaries due to the vulnerabilities in the architecture~\cite{xenakis2006malicious,xenakis2008security} or due to lack of security measures in signaling protocols~\cite{PositiveTechGPRS}. Similarly, exploiting the Domain Network System (DNS), which maps the mobile IP addresses to human-readable hostnames, can also be used as an initial access technique for conducting denial of service attacks on the operator's core network~\cite{tian2017security}. Mobile botnets comprising compromised UEs can also be the initial access for IP-based attacks that result in traffic outage of mobile networks~\cite{anagnostopoulos2016new}. 

The ``\textbf{insider attacks and human errors}'' technique involve the intentional attacks and unintentional mistakes from human insiders with access to any component of the mobile communication ecosystem. Such human insiders could be the employees of operators with evil intent or lack of security knowledge, former employees that can still access the network or whistleblowers. They can exist anywhere from the customer service front desks~\cite{SIMSwapRecruit} to the OSM network that has access to almost all critical nodes of the mobile communication~\cite{bhorkar2017security}. Human errors (including misconfigurations) and the help of insiders bring various threats that forfeit any technical security measures that exist. Hence, they are often leveraged by adversaries as the initial access techniques. Insider threats evolve into a separate category, and it can further be studied with human-centric models that are specifically designed for this purpose\cite{nurse2014understanding}, which is outside the scope of our work.


\subsection{Persistence}
\label{subsec:persistence}
This tactical category represents the group of adversarial techniques for retaining the foothold gained on the target system through the initial access. While it may be sufficient for an adversary to have one-off access to launch an attack, more advanced attacks may require maintaining continuous access, \eg, to complete multi-step attack procedures or just to control the target for a prolonged period. Initial access (\eg, gained through valid credentials) may provide continuous access to the adversary until it is changed. However, the adversary may also want to retain access to the system to avoid interruptions like in the case of system restart, configuration or credential changes.

An adversary retains control over an infected SIM card or UE hardware (\eg, in case of supply chain bugs) as long as it takes the victim to replaces them. Unlike software vulnerabilities, these vulnerabilities cannot be eradicated with a security patch. Malware installed in the case of mobile bots or through the malicious app will give persistent access for as long as they remain undetected by mobile anti-virus solutions or by networks. Likewise, adversaries retain control over the infected network nodes for as long as it takes the operators to detect and patch them.
While there are tools for anomaly detection of traffic emerging from a compromised nodes, routine forensic analysis are rarely done unless there are known security issues. In case of initial access techniques that rely on radio access, the adversary persists its control as long as the victim UE is connected to the spoofed radio network. 

Similar to UE- and network-based malware, it is acceptable to assume that human insiders retain access to the initial foothold for as long as they are undetected. There are various methods and software for insider threat management to detect the presence of human insiders~\cite{sanzgiri2016classification}. However, their effectiveness remains opaque to the general public. A human insider adversary may try to hide its presence from the protection mechanism by opening up covert channels on the point of initial access. Such covert channels could be anything from opening a network port to installing malicious script or remote management software for accessing the compromised entity even if they lose control over their initial foothold. These kinds of persistence techniques are hard to detect without regular sanity checks of network elements for all sorts of software and human vulnerabilities.


\subsection{Discovery}
\label{subsec:discovery}
The discovery tactics include the techniques used by adversaries to gain more information about the surrounding environment of the initial access point, such as system configurations, open ports, and network of other accessible nodes. In most cases, discovery tactics are used only network-side attacks rather than the attacks on the user-side. Depending on the attack, the discovery tactic may be part of the attack mounting or execution phase. If an adversary can launch different attacks or have multiple objectives, the knowledge gained by discovering the surrounding environment will decide its next steps. For instance, if the adversary discovers a protection mechanism that it is incapable of bypassing, they might decide not to proceed further or simply change their original objective. 

Regular Internet users are restricted by their Internet Service Providers (ISPs) with port filtering, where the user's ability to connect or scan a port or IP outside of their normal Internal activities is restricted. ISPs may also have transparent proxies that can selectively censor and monitor the traffic coming from a user. Such filtering rules are the reason why an adversary cannot inspect mobile communication networks even if they are visible to the public Internet. So, an adversary who has gained access to any of the internal nodes of an operator would use \textbf{``port scanning or sweeping''} techniques to probe servers or hosts with open ports. It is a permitted action within the boundaries of network operators as they are often used by network and system administrators for security audits and network maintenance purposes.
However, once inside the network, the adversaries can use the same techniques to determine the open ports and potential services running behind the ports. Such scanning gives information to the adversary about whether a specific host is active on a network and the possibility of compromising it by exploiting a known vulnerability. 

The network of mobile operators is run as a private network and separated from public Internet space. Network Address Translation (NAT) middleboxes are used for translating private IP and port to a public IP and port in the case of establishing an Internet connection from a UE. Such private networks are part of Autonomous Systems (AS), which are routable networks within the public Internet and assigned to individual mobile operators. The Border Gateway Protocol (BGP) routing protocol which allows the AS of the operators to connect to the Internet using their unique ASN number in BGP configuration. An adversary may want to perform ASN and IP lookups alongside port scanning to map their attack surface as a perimeter within which it can access different targets. There is a wide range of publicly available resources for \textbf{``perimeter mapping''} techniques such as command-line utilities (e.g., \texttt{nmap} and \texttt{whois}), web-based lookup tools and official APIs provided by the Internet registrars that assign the ASNs. 

Due to misconfigurations, critical nodes of mobile operators are sometimes are visible over public networks. There are plenty of public and commercial services that compile from multiple sources and share almost real-time threat information of such publicly exposed IPs of critical infrastructure and services behind such IPs with known vulnerabilities~\cite{tounsi2018survey, li2019reading}. There are dedicated search engines (\eg, \texttt{Shodan}~\cite{matherly2015complete} and \texttt{Censys}~\cite{durumeric2015search}) that gather information about vulnerable devices and networks by performing Internet-wide scanning. Similar information can also be found by using traditional search engines with advanced search options, which is popularly known as \textit{Google dorking}~\cite{toffalini2016google}. While such ``\textbf{threat intelligence gathering}'' techniques are essential for mobile operators for tracking public visibility of their security flaws, the same can also be weaponized by adversaries for finding potential initial access or for discovering targets within operator's network.

All the above-mentioned discovery techniques are commonly used for IP-based networks, and they may not be sufficient for an adversary to discover information about the nodes that do not rely on typical IP protocols. More specifically, the older mobile generations use point codes and Global Titles (GT) for uniquely identifying signaling nodes and for signaling routing from core networks. Although such numeric addresses work very much similar to that of IP addresses, adversaries need different tools for scanning nodes that are interconnected with protocols specific to the mobile communication domain. Domain-specific protocols, such as GTP and SCTP (from the SS7 stack), used within the premise of an operator's network are often unauthenticated because a trusted environment is assumed when these protocols are used. Hence, initial access to the network alone suffices for an adversary to prove its authenticity. In such cases, the adversary can use publicly available tools such as \texttt{GTScan}~\cite{GTScan}, \texttt{SigPloit}~\cite{abdelrazek2018sigploit}, \texttt{SCTPScan}~\cite{SCTPScan} and \texttt{GTPScan}~\cite{GTPScan} for discovering the network that is not accessible via IP-based tools. We collectively refer to the use of any such tools as \textbf{CN-specific scanning} techniques.

All the above techniques allow an adversary to discover network topological information, such as addressing, router and gateway filtering, firewall rules, and addressing-based trust relationships that are useful for the next phases of an attack. However, the process is rigorous, time-consuming, and the adversary is at the stake of being detected. On the other hand, if the adversary is, or assisted by, a human insider with internal technical and business documents of a network operator, it can save time and effort. For instance, if the adversary can get access to the operator's IR.21 related resources, which gives it an easy access to the operator's and its partner operators' critical network assets. Every international mobile operator has to maintain up-to-date information about their own network infrastructure details, interconnection, roaming, and inter-operator billing agreements in a standardized manner. GSMA administers such databases containing IR.21 of all operators, and each operator has access to it. Such sensitive information remains out of reach of a regular person, but an insider with access to it can benefit heavily in discovery tactics. We refer to the use of such sensitive databases as the \textbf{Internal resource search} technique.

Finally, just like how IP endpoints and core network nodes are scanned or mapped, adversaries can also do the same for UEs; We refer to this technique as \textbf{UE knocking}. Example uses of this technique can be seen in the case of a radio link attackers or evil mobile operators who test the presence or absence of an UE (based on the associated IMSI) in a given location. UE knocking can also be used for checking other SIM- and UE-specific parameters (\eg, support for specific mobile generation and cipher suites), which helps the adversary in knowing its target more in detail.

\section{Attack execution phase}
\label{sec:execution}
Attack execution is the second phase of the attack life cycle where the adversary launches the attack, to achieve its main objectives, based on the preparation and information that it has sought from the previous phase. We recognize three tactics in this phase, namely, lateral movement, standard protocol misuse, and defense evasion.

\subsection{Lateral Movement}
\label{subsec:lateralmovement}
Once the adversary discovers the surrounding environment of its foothold on the initial access point, it may also want to extend and infect other points of interest. It could either be part of the adversarial strategy or simply the adversary wants to replicate its attacks on all possible points that are accessible to it. In fact, infecting other accessible points would be much easier or different than gaining initial access. We group techniques through which an adversary moves through the target system into the lateral movement tactic.
Similar to discover tactics, lateral movement is used mostly in network-side attacks rather than the attacks on the user-side.

One of the techniques used by adversaries, specifically evil mobile operators, is to \textbf{exploit roaming agreements}. Let us consider a toy example to understand how this technique works in real-world scenarios. Evil operator $\mathcal{E}$ has roaming and business agreements with operator $\mathcal{A}$. Similarly, $\mathcal{A}$ has agreements with $\mathcal{B}$ and $\mathcal{C}$. Such agreements imply that $\mathcal{E}$ can communicate only with $\mathcal{A}$ and not with $\mathcal{B}$ and $\mathcal{C}$, whereas $\mathcal{A}$ can communicate with all $\mathcal{E}$, $\mathcal{B}$ and $\mathcal{C}$. Any communication between operators without a valid agreement will be blocked by the receiving operators to filter unwanted traffic from rogue and arbitrary operators. However, if $\mathcal{E}$ has sought initial access on the premises of $\mathcal{A}$, it can exploit $\mathcal{A}$'s roaming agreements for lateral movement to communicate with  $\mathcal{B}$ and $\mathcal{C}$, which would be not possible otherwise. 

\textbf{Abusing Inter-working functionalities} is another technique for adversaries to move between networks of different generations laterally. Inter-working functions (3GPP TS 29.305~\cite{TS29305}) enable interoperability between the network of operators during roaming scenarios by translating communication messages from one protocol stack to another, \eg, SS7 messages of 2G/GSM to Diameter messages of 4G communication. Adversaries can abuse such functionalities, by using them as attack translation black boxes, for attacking higher generation networks with less secure lower generation networks~\cite{holtmanns2016user}. Even though such functionalities provide an easy way for operating complex interconnections with roaming partners, the flip side of its usage will undermine the security provided by the higher generations.

Although mobile networks seem very complicated from outside, they are nothing but computing machines running commonly used operating systems, software, and services. Once the adversary has infiltrated into the internals of the network, it has ample opportunities and a much broader attack surface to explore. The adversary can, \eg, conduct privilege escalation and process injection for gaining administrative rights, password cracking of valid user accounts on the nodes, exploit vulnerabilities in databases and file systems, and take advantage of improper configurations of routers and switches. We refer to them as the techniques for \textbf{exploiting platform and service-specific vulnerabilities}. This broad technique category can diverge into entirely different threat models that are outside the scope of this work.


\subsection{Standard Protocols Misuse}
\label{subsec:standardprotocol}
In attacks that involve malware or platform-specific vulnerabilities, the adversary can execute a malicious script or override specific configurations. However, adversarial attacks on mobile communication systems are often confined to using standard communication protocols. More specifically, adversaries rely on exploiting their access from a critical node and crafting legitimate messages of standard protocols with evil intent. This is possible because of the continued use of legacy mobile communication protocols built without modern security requirements. The adversaries have to exploit the existence of a feature from such protocols rather than finding any sophisticated vulnerabilities.

One of the examples of standard protocol misuse is in the GSM network, which is one of the oldest but still the most predominantly used network today. With almost no cryptographic security in terms of authentication, confidentiality, and integrity, the Signalling System 7 (SS7) protocol stack used in GSM networks is considered as one of the weakest links. Using SS7 messages as an attack technique by adversaries with access to operator networks has been in a discussion in mobile security communities for over a decade now~\cite{engel2008locating, engel2014ss7, rao2015analysis}. Solutions against these vulnerabilities have also been proposed in various forms, \eg, using secure tunneling~\cite{sengar2005mtpsec, lindskog2008end}, firewalls~\cite{ashdown2001ss7, mehra2019contextual} and machine learning~\cite{jensen2016better}. However, operators are reluctant to deploy the solutions at scale due to operational costs and the burden of its management at a global level. Also, finding hidden features of SS7 and weaponizing them into attack vectors is an active security research theme~\cite{rao2015unblocking}. In this realm, the \textbf{SS7-based attacks} technique continues to be a constant threat until every single GSM network is upgraded. 

Diameter protocol, the successor of SS7 for interconnection in LTE networks, offers better security features than SS7. Despite that, many attacks that rely on SS7 can also be replicated using Diameter, \eg, due to improper deployment of the security features~\cite{mashukov2017diameter, PTSecDiameter, kotte2016detach}. Also, SS7-based attacks can be translated into Diameter attacks using GSM-to-LTE inter-working functions~\cite{holtmanns2016user} even if the adversary has limited knowledge of LTE networks. Since \textbf{Diameter-based attack} techniques slightly differ from their SS7 counterparts, we treat it as a separate technique.

GTP is yet another protocol with no built-in security support \eg, for authentication. Unlike the above-mentioned signaling-based attacks, which originate mostly from an evil roaming partner's or the operator's own premise, the attack surface for GTP is much broader. Adversaries can use \textbf{GTP-based attacks} techniques also from external sources, including the public Internet. In this case, successful attacks result in data interception, Internet service denial, and billing frauds~\cite{PTSecurityGPRS, PositiveTechGPRS}. 

\textbf{DNS-based attack} techniques come from a completely different family of protocols that are not used for signaling in core networks. DNS-based attacks are mainly used in billing frauds as the DNS traffic not counted when metering the data usage of mobile subscribers. Data usage metering is based on TCP packets sent from the mobile endpoints, and the metering starts when the first TCP packet of a data session reaches the boundaries of the core network. An adversary can hide Internet traffic within DNS requests that are part of the data sessions and use the Internet for free-of-charge~\cite{peng2012mobile}. Furthermore, hijacking DNS requests over the radio network allow redirecting them to adversary controlled servers, strip off the encryption, and obtain user's browsing data in plain text~\cite{rupprecht2019breaking}. Such techniques can also be used for generating spam traffic that incurs over-billing discrepancies for the victim subscriber or operator. Similar results can be achieved using GTP-based attacks. However, the adversary has to be within an operator's network.

The radio communication between the UE and mobile network (base station) that takes place before the AKA protocol has no encryption and integrity protection. The UE has to blindly trust the network, which attracts adversaries with radio link capabilities to exploit such trust using \textbf{Pre-AKA attack} techniques. While these techniques are not protocols by themselves, instead, they are the initial phases of standard protocols. Depending on the generation of radio technology, the pre-AKA attack techniques are used (\eg, by IMSI catchers) for collecting both permanent (IMSI and IMEI) and temporary (TMSI) identifiers, for precisely locating a UE within a radio range, for downgrading to lower generations with less security and for denying any mobile service within a radio range~\cite{kune2012location, nohl2014mobile, ShaikIMSICatcher16, golde2013let}. As a well-studied theme, these attacks are addressed in every AKA protocol development with significant improvements. However, even if practically executing such attacks is getting harder, new theoretical loopholes are often caught during formal analysis~\cite{alt2016cryptographic, basin2018formal, cremers2019component, borgaonkar2019new}.

\subsection{Defense Evasion}
\label{subsec:defenseEvasion}
Adversarial techniques used for bypassing protection mechanisms, including evading detection of an adversary's presence, are grouped into the defense evasion tactical category. The techniques under this category could still use techniques from any other tactical groups; however, mainly to subvert the defense mechanism of the system.  

The operating systems, software, and services used on the network nodes are prone to security vulnerabilities and installation of unwanted malware. Although operators conduct routine security audits to track and patch the vulnerabilities or remove the malware from the infected nodes, their effectiveness is not known to the public. Any means by which an adversary can remain undetected from such audits are referred to as the \textbf{security audit camouflage} technique.

Mobile operators employ several defenses in terms of securing their network traffic. For instance, operators maintain a whitelist of IPs and GTs of nodes from their own infrastructure and their partner operators (as agreed in IR 21), and traffic from only these nodes are processed. Similarly, a blacklist is also maintained to control spam due to configuration errors and malicious traffic. Anything from the blacklist is banned from entering the operator's network. Such defense mechanisms may defend against unsolicited traffic from external networks (\eg, from the public Internet and SAN), but it barely serves its purpose in the case of attacks from inter-operator communications. Since most of the communication protocols are unauthenticated in nature, an attacker with knowledge of identifiers of the allowed nodes (\textit{i.e.} gained during the discovery phase) can impersonate their identity. We call it the \textbf{blacklist evasion} technique.

NAT middleboxes are used for separating private networks of mobile operators from public Internet works as the second line of defense. However, studies have shown that the middleboxes deployed by operators are prone to misconfigurations that allow adversaries to infiltrate malicious traffic into mobile networks \eg, by spoofing the IP headers~\cite{wang2011untold}. Some of the other NAT vulnerabilities lie in IPv4-to-IPv6 address mapping logic, which can be exploited by adversaries to exhaust the resources, wipe out the mapping, or to assist with blacklist evasion~\cite{hong2017cellular}. Adversaries use such \textbf{middlebox misconfiguration exploit} techniques to launch denial-of-service or over-billing attacks~\cite{leong2014unveiling}.

A more advanced form of defense against signaling attacks is available in the form of firewalls. Some operators deploy stateful firewalls that are readily available as commercial~\cite{SinchSigFirewall} and open-source products~\cite{SigFW}. While only a small fraction of operators use such firewalls~\cite{SigSecENISA}, they are expected to distinguish legitimate network traffic from malicious ones by tracking the operating state and characteristics of network traffic traversing it. However, adversaries (e.g., evil operators) can exploit the implicit trust between roaming partners as a \textbf{bypass firewall} technique. As a reminder, signaling attacks that are based on SS7, Diameter, and GTP protocols are no different from regular messages exchanged between operators in genuine roaming scenarios~\cite{puzankov2017stealthy}. By hiding attack traffic amongst the massive amount legitimate signaling traffic exchanged between roaming partners, adversaries can always expect it to bypass the firewall protection. Other solutions, such as using machine learning features in firewalls ~\cite{jensen2016better}, are still in their early stages of research. As bypassing techniques would also evolve alongside firewall protection mechanisms, we expect such attack techniques to remain valid in the future. 
%

Most of the SS7-based attacks exploit the flaw in the SMS delivery mechanism for inbound off-network SMS messages where delivering SMS is assumed to be the responsibility of the operator from where the SMS is originated rather than the operator to which the SMS needs to be delivered. In genuine cases, the operator of the SMS's origin requests the operator (i.e., roaming network) where the SMS has to be delivered about the location of the UE roaming in its network. The latter has no option other than to pass the location information to such requests from an external network with no means of authentication. However, such mechanisms can be exploited by a core network attacker who can spoof of an operator to deliver SMS and obtain the location of the UE if the IMSIs of its victims are known. SMS home routing is a defense mechanism~\cite{TR23840}), where an additional SMS router intervenes in external location queries for SMS deliveries, and the roaming network takes the responsibility of delivering the SMS without providing location information to the external entity. Although many operators have implemented SMS home routing solutions, there are no silver bullets. If the SMS routers are incorrectly configured, adversaries can hide SMS delivery location queries within other messages so that the SMS home router fails to process them~\cite{HomeRoutingBypass}. We refer to it as the \textbf{bypass home routing} technique.

Attacks on the radio access networks are well-studied and newer generations are designed to address the weaknesses in previous generations. Usage of weak cryptographic primitives, lack of integrity protection of the radio channels, and one-sided authentication (only from the network) remain as the problem of mostly GSM only radio communication. So, radio link attackers use \textbf{downgrading} as an attack technique to block service over newer generations and accept to serve only in the GSM radio network. The downgrading technique works similarly in the core network, where the adversary accepts to serve only in SS7-based signaling instead of Diameter-based signaling. Using interworking functions for inter-generation communication translation could make the downgrading attacks much easier~\cite{holtmanns2016user}.

\textbf{Redirection} technique is a variant of the downgrading technique, where an adversary forcefully routes the traffic through networks or components that are under its control. By redirecting traffic to an unsafe network, the adversary can intercept mobile communication (\eg, calls and SMS) on the RAN part~\cite{zhang2016lte}. Redirection attacks on the core network result in not only communication interception, but also in billing discrepancies, as an adversary can route the calls of a mobile user from its home network through a foreign network on a higher call rate~\cite{engel2014ss7}. 

Protection on the UE is mainly available in the form of antivirus apps as a defense against viruses and malware that steals sensitive information (\eg, banking credentials and user passwords) or track user activities. Simple visual cues on UE (such as notifications) could also be a protection mechanism by itself.
Unfortunately, mobile network-based attacks cannot be detected or defended effectively from UE's side by traditional antivirus apps, and such attacks do not trigger any visual signs. Although there are attempts for defending against radio link attacks~\cite{dabrowski2014imsi,dabrowski2016messenger,borgaonkar2014understanding, van2015detecting},  including city-wide studies to detect IMSI catchers~\cite{li2017fbs,ney2017seaglass}, their effectiveness is still under debate~\cite{borgaonkar2017white}. Similarly, there are recent attempts to detect signaling attacks using distance bounding protocol run from a UE~\cite{peeters2018sonar}. However, such solutions are still in the research phase, and their effectiveness on a large scale is still untested. To this end, the absence of robust detection and defense mechanisms on the UE is, in fact, an evasion mechanism for an adversary. We refer to them as \textbf{UE protection evasion} techniques. 

\section{Attack results phase}
\label{sec:resultphase}
Attack results are the last phase of the attack life cycle, where the adversary achieves his main objectives. The tactics in this phase imply the end result, in terms of collection of information that is of utmost interest to the adversary and final impact of an attack.
%

\subsection{Collection}
\label{subsec:collection}
The collection tactics represents sensitive information gathered or stolen by an adversary by achieving any of its tactical objectives during attack mounting or execution phase. 

Stealing legitimate \textbf{admin credentials} for critical nodes is beneficial for the adversary to increase its chances of persistence to the target or masquerade its activities. There are various well-known methods for exploiting platform and service-specific vulnerabilities (\eg, key-logging and brute-forcing) by which adversaries can collect such information~\cite{AttckCredentialAccess}. Mobile communication systems offer various other sensitive information as well. 


\textbf{User-specific identifiers} such as IMSI and IMEI are an indicator for who owns UE with a specific subscription and where a UE is located physically. Since mobile users always keep their mobile phones physically near them, an adversary with the knowledge of these permanent identifiers will be able to determine whether or not a user is in a specific location. On the other hand, temporary identifiers (\eg, TMSI and GUTI) are used to reduce the usage of permanent identifiers like IMSI over radio channels. Although the temporary identifiers are supposed to change frequently and expected to live for a short period, research has shown that it is not the case~\cite{arapinis2014privacy, hong2018guti}. Reuse of these temporary identifiers forfeits its benefits and yields to mapping them back to IMSI. Hence, the collection of temporary identifiers is equally valuable as that of the permanent ones. We consider encryption keys that reside in the SIM cards and the MSISDN (phone numbers) also as user-specific identifiers as they are equally attractive to adversaries. Unlike the rest of the identifiers, MSISDN is public by nature. However, the collection of arbitrary MSISDNs from a specific location~\cite{yu2019lte} makes it potentially sensitive like the rest of the identifiers. Besides the obvious privacy concern of sensitive information to unintended entities, the user-specific identifiers also form a crucial part of adversarial tactics, for example, for the successful use of standard protocols (~\ref{subsec:standardprotocol}) in the core network communication. 


The acquisition of user-specific identifiers suffices in case, say if the adversary's motive is to obtain the location of a specific user. However, the adversary can collect several types of \textbf{user-specific data} if he has a much broader motive. Such data include, \eg, the content of SMS and calls, location dumps from base stations, call and billing records, and browsing-related data (such as DNS queries and unencrypted browsing sessions).

Adversaries aim to collect \textbf{network-specific identifiers} such as GTs and IPs of critical nodes and Tunnel Endpoint Identifier (TEID) of GTP tunnels from operators' networks. Adversaries may also be interested in \textbf{network-specific data} that are obtained mainly during the execution of discovery tactics. Such data includes, \eg, the network topology, the trust relationship between different nodes, routing metadata, and sensitive documents. 


\subsection{Impacts}
This tactical category contains the main objectives achieved by the adversary by following a series of techniques that belong to all previous tactical categories.

Some degree of user location tracking is required for the fundamental working of mobile network technologies (\eg, for continuous handovers), where a UE regularly discloses its location to the mobile networks in legitimate scenarios. The same can be exploited by different kinds of adversaries, from almost every network subsystem, for conducting \textbf{location tracking} attacks. For example, radio link attackers can use fake base stations either to check whether UEs are present in a given location or to obtain precise coordinates of the UE ~\cite{ kune2012location, park2019anatomy, ShaikIMSICatcher16, jover2016lte}. Similarly, adversaries with access to core network can misuse signaling protocols (\eg, SS7~\cite{engel2008locating,engel2014ss7} and Diameter~\cite{holtmanns2016user,rao2016privacy}) or exploit vulnerabilities in the signaling plane~\cite{roth2017location} for obtaining location information. IP-based attack vectors from the IMS domain (\eg, over VoLTE) also yields similar results~\cite{kim2015tracking}. The precision of location obtained, however, varies --- from a few meters to tracking area of an MME to a country-wide service area of an operator --- in accordance with the adversary's capabilities and objectives. Although location tracking attacks directly affect the privacy of the mobile users, their extended implications include using the obtained location for spoofing against core networks~\cite{hussain2018lteinspector}, denying service to users, and incurring financial loss to operators. We believe that the location tracking attacks are used as a preparatory step for enabling other attacks that we discuss further in this section. 

Voice calls and SMS are two of the native applications of mobile communication systems that lack features (such as end-to-end encryption) that can be achieved with modern secure communication protocols.
 Although there have been proposals (\eg, for SMS encryption), they have been disregarded by standardization bodies and operators due to their deployment complexity~\cite{lo2008smssec,saxena2014easysms}. Currently, only the radio channel between UE and base stations is protected with encryption, and everything beyond that (\ie, inside core network) is transported in plain text. Therefore, the security of calls and SMS relies on the strength of encryption for over-the-air radio communication and blind trust on the operators for traversal through the core network. Unfortunately, both of these are vulnerable to \textbf{calls eavesdropping} and \textbf{SMS interception} attacks. 
 
On the radio network side, adversaries can use fake base stations to force a UE to transmit all its communication without encryption in some cases and intercept the non-encrypted transmission of voice calls and SMS. However, it is not always possible. Adversaries often rely on weaker encryption schemes to achieve passive interception. For example, many cryptanalysis~\cite{wagner1997cryptanalysis,biryukov2000real, biham2005related, barkan2008instant,dunkelman2010practical} and rainbow table-based attacks~\cite{nohl2011defending, papantonakis2013fast} have shown that encryption schemes used in GSM radio communication can be broken. Since this is mostly the problem of GSM networks, adversaries use downgrading technique (refer to Section~\ref{subsec:defenseEvasion}) from a higher generation network, passively record the encrypted communication so that it can be decrypted later.
 
Attacks from the core network, on the other hand, allow an adversary to intercept and eavesdrop calls and SMS actively. More specifically, regular call setup and SMS delivery workflows during legitimate roaming scenarios give away control to route the calls and SMS to external entities (roaming partners). Unfortunately, there is no means of verifying authorization of the request from external entities during such scenarios. An adversary can abuse it to redirect the calls and SMS to an unsafe network (or node) that it controls. Since communication within the core network occurs without confidentiality protection (\ie, no encryption), the adversaries can intercept them easily. Such attacks are peculiar artifacts of SS7-based attacks in GSM networks. Due to a lack of authentication and authorization of signaling messages, adversaries can either impersonate legitimate core network nodes or manipulate the mobile subscriber profiles to route the traffic towards the nodes under their control~\cite{engel2014ss7,nohl2014mobile,puzankov2014intercept}. Recent studies have shown that similar attacks are also possible in LTE networks using Diameter-based attack vectors~\cite{holtmanns2017subscriber,holtmanns2017sms}.



\textbf{Data (Internet traffic) interception} attacks work differ from call and SMS interception. An adversary can intercept or modify the content by stripping off the encryption on the radio link layer or injecting messages via signaling protocols (GTP in this case). However, it will be limited only to avail free internet at the cost of someone else using redirection techniques. To tamper with the actual content of Internet traffic, adversaries need a more sophisticated approach, especially for modern LTE networks, which overcomes many security issues of previous generations. Here, we focus mainly on data interception in LTE networks from man-in-the-middle adversaries.

As a reminder, mobile communication comprises user plane, which contains the actual content (e.g., websites visited), and control plane, which involves radio and signaling control messages on how the user traffic should be communicated in the network. The control plane in LTE radio transmission is both encrypted and integrity protected, which forbids an adversary from controlling how the user traffic should be transmitted. However, the user plane has only encryption, and this leads to various issues. Firstly, this setup allows passive website fingerprinting from browsing metadata, where a radio link attacker can learn a user's accessed website~\cite{kohls2019lost} or perform chosen plain-text attacks on the encrypted user plane traffic~\cite{rupprecht2019breaking}. Secondly, adversaries can impersonate the victim towards network due to network misconfiguration and implementation errors~\cite{chlosta2019lte}. Such errors, \eg, force the UE to accept \textit{null encryption}, which result in unencrypted user plane traffic to be intercepted by passive adversaries~\cite{rupprecht2016putting}. Similarly, fuzzing of exception handling has shown that UE can be tricked to communicate unprotected user data through an adversary-controlled rogue LTE radio networks~\cite{kim2019touching}. Finally, DNS traffic remains alterable in LTE RAN, which allows the adversary to perform DNS spoofing and data redirection attacks to intercept the user traffic in unencrypted format~\cite{rupprecht2019breaking}. Most of these data interception attacks work mostly under controlled lab setup, and they are not observed in the wild. In most of the cases, we believe that application-layer security (HTTPs) offers protection to user's internet traffic.

\textbf{Billing frauds} refer to various types of attacks where an adversary causes financial discrepancies for operators. We suggest the readers to refer to the work by Sahin et al. ~\cite{sahin2017sok} for an extensive overview and classification of billing frauds. In most cases, the adversary's objective is to avail services offered by operators for free of charge. The first kind of billing fraud includes voice calls and SMS. Adversaries can generate spoofed and spam SMS and calls, \eg, by signaling traffic manipulation~\cite{engel2014ss7, nohl2014mobile} and fake base stations ~\cite{FakeBSBasedSMSSpam}. Similarly, the Internet-based services for calls and SMS (including SIP, VoIP and VoLTE) have no means to restrict or verify the authenticity of the origin of the communication, which results in a wide range of billing frauds~\cite{tu2016new,chalakkal2017practical,collier2013hacking, zhang2010billing,li2015insecurity,kim2015breaking}. The second kind of billing fraud includes mobile Internet services. Adversaries here (which also includes mobile users) alter the mobile Internet traffic, \eg, by spoofing mobile IP headers~\cite{peng2014real,wang2011untold} or by manipulating TCP and DNS requests~\cite{go2014gaining,go2013towards} and cause financial discrepancies in an operator's accounting of data usage. 




We segregate the denial of service attacks into two categories. The first one is the \textbf{Denial of Service (DoS) -- Network} attacks which relies on creating signaling havoc in specific nodes of operators by repeatedly triggering resource allocation or revocation requests. Such attacks are capable of exhausting powerful core network nodes such as the HLR. Most of the signaling DoS attacks are targeted towards exhausting the the RAN by abusing radio channel allocation requests~\cite{lee2009detection,ricciato2010review,kambourakis2011attacks,bassil2013effects,bassil2012signaling,golde2013let}. Another major category of DoS attacks arise from SMS-capable interfaces from IMS domain and from public Internet where operators have least control~\cite{croft2007silent,enck2005exploiting, traynor2008mitigating,tu2016new}. Similarly, cellular botnets ~\cite{traynor2009cellular,khosroshahy2013botnets} (including SMS-botnets~\cite{geng2012design,zeng2012design}) can launch DoS attacks from a large army of compromised UEs. 
  
The second category is of \textbf{Denial of Service -- User} attacks. Although they share commonalities with the DoS -- Network techniques, they are targeted towards denying service to mobile users. The most popular DoS attacks on users are in the form of radio signal jamming~\cite{lichtman2013vulnerability, lichtman2016lte, xiao2013analysis, jover2013security, aziz2014resilience}. Radio link adversaries use open-source tools for building such jammers and forbid users from joining mobile networks~\cite{rao2017lte}. Such DoS attacks on the radio interface are often seen in conjunction with other types of interception and location tracking attacks~\cite{shaik2015practical}. Another type of DoS -- User attack emerges from core network adversaries who can alter SS7 and Diameter protocol messages~\cite{engel2014ss7, kotte2016detach}. Unlike jamming, the adversary here modifies subscription profile data of targeted mobile users while sending location update requests in inter-operator roaming scenarios. More specifically, the adversary deletes or changes location information of a roaming user in the subscription profile and convinces the serving network to deny services, such as calls or SMS to victim users.

\textbf{Identity-based attacks} involve attack techniques using user- and network-specific identifiers (refer to section~\ref{subsec:collection}). Identity-based attacks cause harm to the privacy of mobile users and produce fraudulent traffic that incurs a financial loss to operators. In most cases, identity-based attacks are used in \textit{impersonation}, where an adversary impersonates a legitimate mobile user to the core network without possessing appropriate credentials, for example, to avail free mobile services~\cite{rupprecht20imp4gt,hussain2018lteinspector}. Most of the signaling attacks that use SS7 are also fall into this category. In other cases, identity-based attacks involve \textit{identity mapping}, where the adversaries map temporary identifiers (\eg, TMSI and GUTI) to permanent identifiers (\eg, IMSI or MSISDN)~\cite{arapinis2014privacy,hong2018guti,rupprecht2018security,yu2019lte}. In rare cases, the IMSI can further be mapped to social media identities~\cite{shaik2015practical}.

\section{Use cases of the threat model}
\label{sec:usecases}

In this section, we present two use cases of the Bhadra threat modeling framework with the help of concrete examples. More specifically, we demonstrate how to use Bhadra for modeling individual attacks independently and for modeling and comparing multiple attacks all together. 

\subsection{Modeling individual attacks}
\label{subsec:casestudy-newones}

As exemplars of recent attacks, we consider \sj~\cite{SIMjacker} and \mt~\cite{MessageTap} for our case studies. We chose them because neither of them has been studied in any of the literature that we considered for building our threat model. As the recently discovered attacks, they both give us fresh perspectives on the sophistication and evolution of adversarial tactics and techniques of state-of-the-art attacks on mobile communication systems. These attacks are believed to be conducted by sophisticated attacker groups. The existence of adversary groups allows us to look at the adversarial behavior through the lens of our framework. Furthermore, given that these attacks span over most of the tactical categories and multiple techniques in each tactic, they are ideal for us to show the use case of our framework to model individual attacks independently.

\subsubsection{\textbf{Case study 1: \sj}}
\label{subsubsec:simjacker}
\hfill
\newline
\sj was publicly disclosed in September 2019 by AdaptiveMobile Security during their lookout for identifying unexpected behavior and previously undetected suspicious activity in mobile communication networks~\cite{simjackerdisclosed}. \sj is a large scale espionage attack on mobile users in multiple countries, presumably from a competent adversary group on behalf of a nation-state actor. In its entirety, the adversary here exploits vulnerabilities in SIM cards with SIM alliance Toolkit (S@T) browser to execute SIM-specific functionalities without the knowledge of the targeted mobile user. The adversary sends the attack payload in the form of a specially formatted binary SMS to the victim's phone, either through a regular UE, VAS provider, or SS7 protocol. A successful attack yields the location and IMEI of the user to the adversary as a reply SMS without the notice of the mobile user. 

S@T browser specifications were developed by the SIM Alliance~\cite{SATBrowserguidelines}. It allows running applications in the SIM card using commands from over the air (OTA) SMS. Unlike the regular SMS for sending text messages between mobile users, OTA SMS is a special form of binary SMS that issues commands to be executed on the SIM card as per 3GPP TS 31.111~\cite{TS31111}. Typically, such binary SMS are sent from mobile operators to their subscribers to configure the SIM Card to initiate various value-added services. Since such commands are expected only from legitimate network operators, authentication to execute the commands is not implemented. Commands that are specific to S@T browser can, \eg, power off the SIM card, collect UE-specific information (such as the location, IMEI, and battery status) and send them back via mobile services (\ie, SMS, MMS, USSD), launch a web browser or make phone calls. 

\begin{figure*}
\centering
\begin{subfigure}{\textwidth}
   \includegraphics[width=\textwidth]{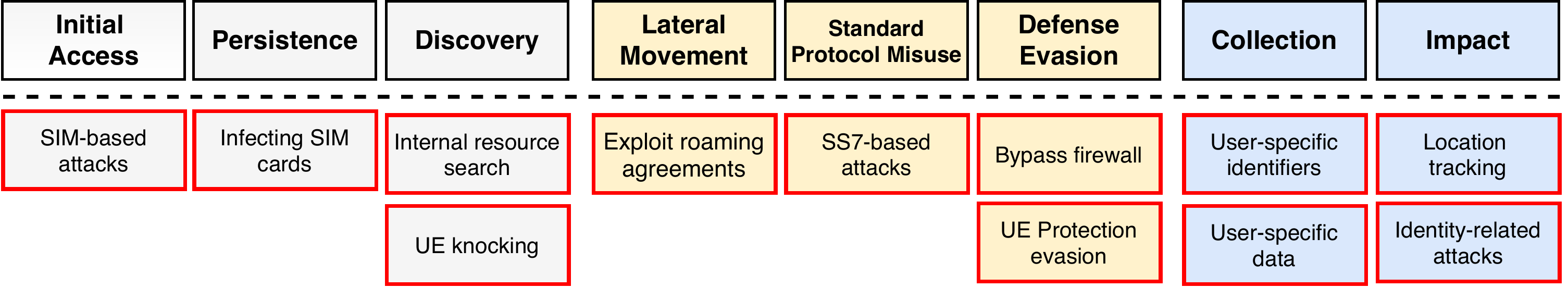}
   \caption{Simjacker attack modeling}
   \label{fig:SimjackerModel} 
\end{subfigure}

\bigskip 
\bigskip 

\begin{subfigure}{\textwidth}
   \includegraphics[width=\linewidth]{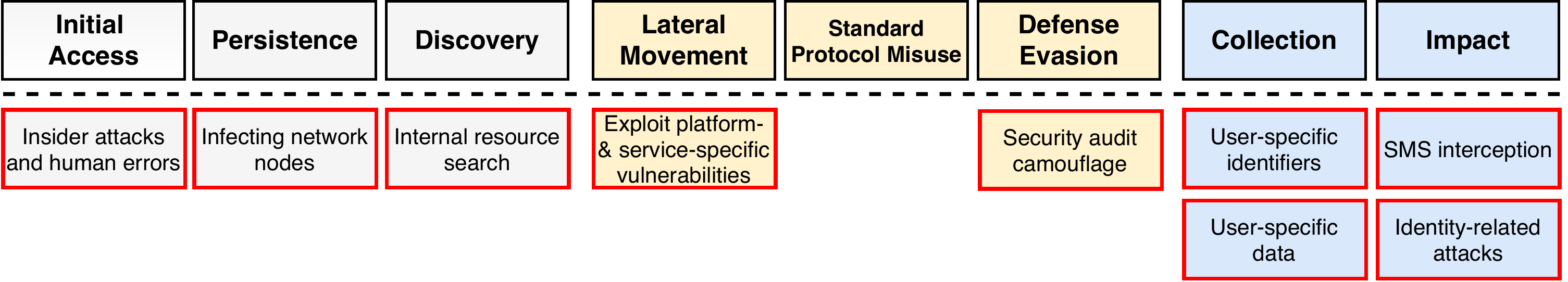}
   \caption{MESSAGETAP attack modeling}
   \label{fig:MessageTapModel}
\end{subfigure}

\caption{Modeling recent attacks using Bhadra framework (only relevant techniques are shown)}
\end{figure*}

We now model the \sj attack with the Bhadra framework (refer to Figure~\ref{fig:SimjackerModel}) and explain the attack procedures in detail. 

\boldtitle{Initial access} --- Although the attack originates elsewhere, the first point of access for the adversary is the SIM cards with S@T browser functionalities. Therefore,  we tag the ``\textit{SIM-based attacks}'' technique for the initial access tactical category.

\boldtitle{Persistence} --- The S@T browser is embedded to the SIM card, and its functionalities are hard-coded. So, the adversary can exploit the vulnerabilities as long as the SIM card is replaced with one without the S@T browser. Therefore, we tag the ``\textit{infecting SIM cards}'' technique for the persistence tactical category.

\boldtitle{Discovery} --- To know the presence of the S@T browser, the adversary has two potential options. Firstly, by sending test messages to the target UE (\ie ``\textit{UE knocking}'' technique). Secondly, from ``\textit{internal resource search}'', where the mobile operators or their card manufacturers may have the list of SIM cards issued with S@T browsers. 

\boldtitle{Lateral movement} ---  The initial target of the adversary was Mexican mobile subscribers; however, eventually, Colombian and Peruvian subscribers also became prey. Also, the analysis reveals the reliance on SS7-based attack vectors~\cite{SIMjacker}. Based on these, we assume that the adversary uses the ``\textit{exploit roaming agreements}'' technique to move laterally from the original victim operator to its partner operators.

\boldtitle{Standard protocol misuse} --- As per the analysis of \sj attack activities on a global scale, a close coordination of ``\textit{SS7-based attacks}'' were seen carrying the SMS payload~\cite{SIMjacker}. Although most of the attack SMS was originated from actual UEs, a significant fraction was also originated from SS7 addresses (\ie, from GTs of other mobile operators) if the UE-originated SMS were not delivered to the victim mobile subscriber. 

\boldtitle{Defense evasion} --- Exploiting binary SMS is not novel by itself. Such attacks are demonstrated in information security conferences ~\cite{nohl2013rooting, alecu2013sms} and also used in the wild by nation-state actors~\cite{inside2014documents,marczak2016million}. Due to these disclosures, operators have implemented blocking on the ability of SIM card to process binary SMS, as per 3GPP TS 23.048~\cite{TS23048}. However, the \sj attack circumvents any such defense on the UE side with sophisticated SMS packet encoding. It is observed that modification of SMS headers, non-standard binary SMS formats, and multi-parted packet creations are used while encoding the SMS packets to route and processed by SIM cards. Hence, we tag ''\textit{UE  protection evasion}'' as one of the defense evasion techniques. 

For evading defense during network traversal of protocol messages carrying the SMS payload, the adversary uses the ``\textit{bypass firewall}'' technique. More specifically, for the SS7-based variants, the adversary impersonates (by using valid GTs) the MSC/MME and SMS center (SMSC) from the targeted operator's partners. Similarly, the payload was also sent from valid VAS providers directly to the SMSC of the targeted operator. In both cases, the impersonation of trusted sources helps the adversary to avoid filtering by firewall rules and to transport the payload in and out of the targeted operator's network.

\boldtitle{Collection} --- The adversary's binary SMS instructs the SIM card on the target UE to collect its current serving cell-ID (``\textit{user-specific data}'') and IMEI of the UE (``\textit{user-specific identifier}''). This collected information will be sent back to the adversary, also, in the form of a binary SMS.

\boldtitle{Impact} --- Although attacks that exploit the S@T browser can perform various dangerous functionalities, \sj attacks have targeted primarily only on large scale collection of cell-ID and IMEI of UEs. We therefore tag ``\textit{location tracking}'' and ``\textit{identity-based attacks}'' respectively for the adversary's impact techniques.

\subsubsection{\textbf{Case study 2: MessageTap}}
\hfill
\newline
\mt was publicly disclosed in October 2019 by the FireEye threat research team during their investigation of a network provider~\cite{MessageTap}. Based on the evidence, it is attributed to APT41 Chinese APT group~\cite{apt41} in support of state-sponsored espionage campaign by the Chinese government. \mt is a malware that sits on the SMSC servers of the operators and logs SMS content and contact network of specific individuals (based on their IMSIs and phone numbers) that are of geopolitical interest for the Chinese intelligence. SMSC is a Linux server that is mainly responsible for routing SMS messages to intended mobile subscribers. However, if the recipients are not online (\ie, not attached to the mobile network), SMSC stores the SMS for them until the subscriber is available to deliver. 

\mt is a Linux Executable and Linkable Format (ELF) malware specially crafted for SMSC, and its working mechanism is as follows. \mt initiated on the SMSC by an installation script. The script also involves two data files, one with target IMSIs and phone numbers and the other with a list of keywords to match. Both these files are erased from disk once they read and loaded into memory. \mt then starts to monitor all network connections to and from the SMSC, parsing, and extracting the SMS data from the network traffic based on the content of data files loaded into memory. More specifically, the malware looks for specific IMSIs, phone numbers, and keywords from the data files in the SMS message flow. If a match is found, the IMSI, source and destination phone numbers, and content of the SMS message are stored in separate files to be sent to the adversary. 

We now model this attack with the Bhadra framework (refer to Figure~\ref{fig:MessageTapModel}) and explain the procedures in detail.

\boldtitle{Initial access} ---  SMSc is an operator-specific entity, and installing a malicious script on it is unlikely done by arbitrary adversaries. Although there are no clear evidence of how the script was installed on the SMSC, given the involvement of nation-state actors, we speculate that it is deliberately done by insiders from the mobile operators. We, therefore, tag the ``\textit{insider attacks and human errors}'' technique for the initial access tactic.

\boldtitle{Persistence} --- In terms of persistence, the adversary's technique falls into the category of ``\textit{infecting network nodes}''. Once the malware is installed on the SMSC, the adversary's control over it persists as long as the malware is not removed or at least its functionalities are restricted. 

\boldtitle{Discovery} --- We tag ``\textit{internal resource search}'' for the adversary's discovery technique for discovering the details of SMSC and the possibilities of installing a script for logging SMS messages. Although there is no clear evidence on how the adversary discovered it, we suspect the involvement of insiders leaking internal resources about such details.

\boldtitle{Lateral movement} ---  The malware installation probably exploits improper access control or authorization on the SMSC server. While we do not know the exact reason, like any malware, in general, we believe that the \mt uses the ``\textit{exploit platform and service-specific vulnerabilities}'' technique for lateral movement. Since everything occurs on the SMSC, there is only a slight difference in lateral movement tactic from the initial access. The latter refers to an insider having bare minimal access to the SMSC server, whereas the former involved leveraging such access to install malware.

\boldtitle{Standard protocol misuse} --- Even though SMS communication is part of the standard protocol (SS7 in this case), the protocol itself has no role in the attack execution. So, we leave the ``\textit{Standard protocol misuse}'' blank in our model. 
\begin{figure*}
  \includegraphics[width=\textwidth]{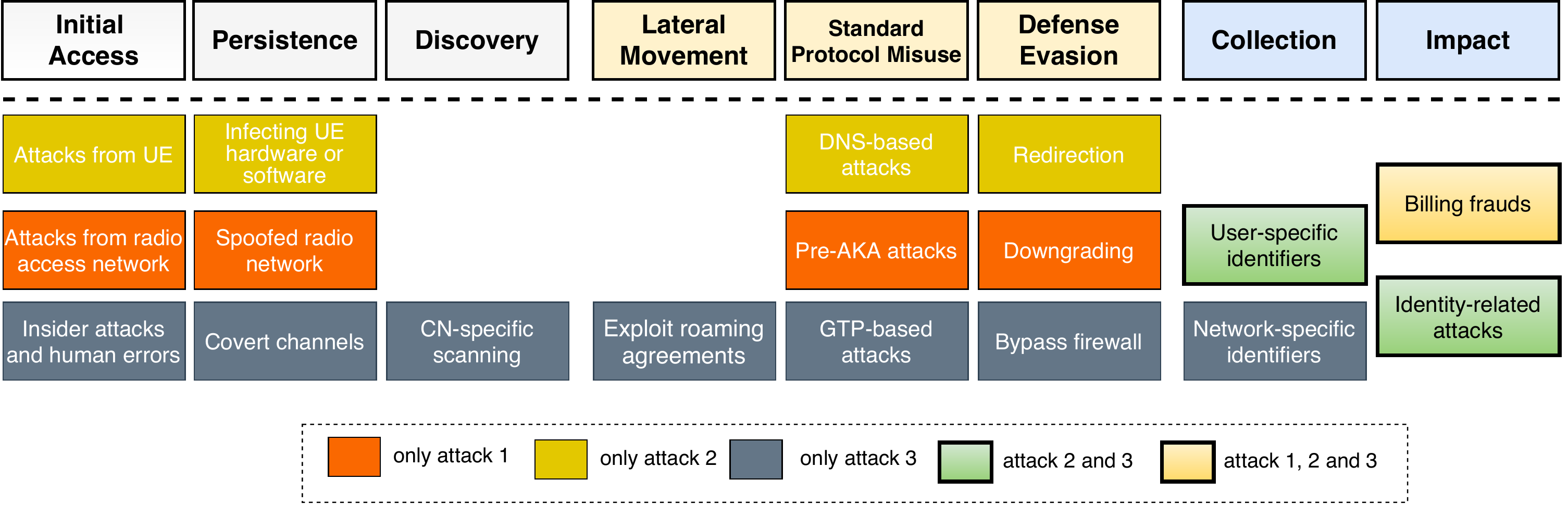}
  \caption{Comparing billing fraud attacks using Bhadra framework (only relevant techniques are shown)}
  \label{fig:ComparisonMatrix}
\end{figure*}

\boldtitle{Defense evasion} --- As mentioned before, after the installation, the malware starts capturing the network traffic and parses all layers of the protocol. This action typically requires elevated access rights on the network interface or causes discrepancies (\eg, delay) in the traffic. The victim operator detected neither the installation nor the network interception during their routine security audits. Hence, we tag the ``\textit{security audit camouflage}'' as the defense evasion technique.

\boldtitle{Collection} --- The \mt collects the content of the SMS message based on a match with pre-defined keywords, IMSIs and phone numbers contained in the SMS communication of specific individuals. Hence, we tag ``\textit{user-specific data}'' and ``\textit{user-specific identifiers}'' respectively as the collection techniques.

\boldtitle{Impact} --- The main objective of the adversary here is to log the contents of the SMS. So, we first tag the \textit{SMS interception} technique for the impact tactics. Secondly, the malware also builds a contact network of specific individuals by collecting IMSIs and phone numbers associated with the SMS communication. We therefore tag \textit{identity-based attacks} also as an impact.

\subsection{Comparing multiple attacks}
\label{subsec:casestudy-comparing}

For the comparison, we pick three attacks from the literature that result in billing fraud. The only commonality between these attacks is the adversary's objective to use data services of an operator for free of cost or at the expense of others. To do so, the adversary obtains an IP address on behalf of a legitimate mobile user and tricks the charging system into charging the victim subscriber (or the operator in some cases) for all traffic used by the adversary. All three cases will incur direct financial losses and accounting discrepancies to the mobile operator. We do not go through individual tactics one by one, as we did in the previous section. Instead, we briefly describe all three attacks and fast-forward to model and compare them. The result is presented in Figure~\ref{fig:ComparisonMatrix}.

\boldtitle{Attack 1} --- The first billing attack we consider is presented by Peng et al.\cite{peng2012mobile} in the form of two types of frauds, namely, \textit{toll-free data service} and \textit{stealth spam} attacks. Even though both allow the adversary to enjoy mobile data services at the expense of someone else, we consider only the former attack. In the \textit{toll-free data service} attack, the adversary is a mobile subscriber who is capable of crafting malicious data packets from the UE.
The attack is based on the observation that mobile operators transport DNS traffic without charging or limiting its usage. So, the adversary has two options to exploit this. The first one is to redirect TCP or UDP traffic from DNS port number 53, as many operators do not restrict the traffic from port 53, which is expected to send only DNS traffic. However, if non-DNS traffic from port 53 is restricted, the adversary can use the second option of redirecting using fake DNS requests. The actual data resides inside the malformed DNS request, and it goes undetected by the operator. In any case, the adversary will be able to enjoy free data services. 

\boldtitle{Attack 2} --- The second billing fraud attack that we consider is presented by Chlosta et al.~\cite{chlosta2019lte}. The adversary here is a radio access attacker who lures mobile subscribers in his vicinity to connect to its spoofed radio access. The adversary acts as a man-in-the-middle between a benign UE and the eNodeB to impersonate the UE to the operator network and vice versa. As per LTE specifications, mutual authentication between UE and eNodeB is ensured with AKA protocol and the subsequent integrity protection of the control plane. However, operator-specific implementations often fall back to insecure scenarios, such as allowing null-integrity and null-encryption. The adversary collects the IMSI from the UE and enforces the selection of null algorithms by the benign network using the downgrading technique. He then impersonates the benign UE to the operator to obtain an IP. All following data services are billed towards the victim as the Internet connections are associated with its IMSI. In summary, the adversary obtains IMSI of the victim and free Internet usage that is billed towards the victim as well. 

\boldtitle{Attack 3} -- The final attack that we consider is presented by Positive Technologies~\cite{PositiveTechGPRS}. The adversary here could be a human insider with core network access. 
The adversary exploits the fact that IP addresses of the UE that sends a request to the operator's core network entities (for data service access purposes) are not thoroughly verified, especially if they come as part of a core network protocol message. The adversary has to spoof the subscriber’s identity (IMSI) and uses GTP-based techniques to send ``\texttt{Create Session Request}'', a GTP control plane service message to the PGW (refer Figure~\ref{fig:topology}) to gain Internet access at the cost of someone else. If the spoofed IMSI belongs to a legitimate subscriber, the charging system of the partner operator will charge that subscriber for all traffic used by the adversary. If the IMSI is bogus, the partner operator will have to bear the cost of data services used by the adversary.  Unlike attacks 1 and 2, even though the attacker is an insider, he has to perform additional steps to figure out the correct address of the PGW of the partner operator through CN-specific scanning, exploiting inter-operator roaming agreements and bypassing firewalls that forbid from sending malicious GTP traffic.  

As we can see in Figure~\ref{fig:ComparisonMatrix}, Bhadra framework is used for comparing multiple attacks where the main objective of the adversary is to enjoy data services for free, which in turn results in billing frauds. Though we followed the same steps for modeling individual attacks as we did in Section~\ref{subsec:casestudy-newones}, we color-coded them so that common techniques between the attacks can be highlighted with an overlapping color. Although the objective is the same in all three attacks, as we can see from modeling comparison, the adversarial paths to achieve them differs extensively from each other. Of course, the adversarial paths depend very much on the adversary's capabilities rather than the objective itself. One might assume that for a human insider (attack 3), it is much easier to conduct billing frauds. However, on the contrary, such adversaries have to perform the additional steps of discovery, and lateral movement techniques, which the adversary closer to user endpoints (from attack 1 and 2) can skip. On the other hand, attack 1 and 2 involve building custom attack tools to tinker with radio hardware or mobile application-level programming, which requires more technical knowledge, unlike using operational network functionalities that the human insider can use in attack 3. This way, we believe that the additional steps taken by attacker 3 compensate in terms of the difficulty of the attack. 

While what we have compared and discussed is just an example, one could infer various other things by comparing attacks that have more commonalities. One specific thing that is more interesting in all cases is how the adversarial paths, which depicts the behavioral aspects of the attacks, differ from each other despite their commonalities.





\section{Discussion}
\label{sec:discussion}

Initial generations of mobile communications are neither based on strong cryptographic primitives nor built with modern security needs. Instead, it started from forbidding access to gradual additions of security features to various non-operators as the need for various kinds of services expanded. Lack of stronger security building blocks in the initial generations have bothered till the current ones. 5G seems to end this trend by starting everything from scratch. For example, there are proposals for the use of public-key infrastructure~\cite{hussain2019insecure} or formal verification of the protocols~\cite{borgaonkar2019new, hussain20195greasoner, cremers2019component}, all of these which have not been done before. We believe that our framework fits into this growing trend of learning from past mistakes and trying out methods that are not previously practiced. 

At this point, we believe that our framework is just the first iteration, as this is the first of its kind. It can grow mature in the future with more contributions from the mobile security community. The modular nature of the framework makes it easier to add more techniques to existing tactical categories or even add additional tactics if there are a sufficient number of techniques, that too, without disturbing the rest of the framework. 

Furthermore, the matrix representation of the framework makes it an excellent candidate to be used for quantitative analytic purposes. More specifically, by modeling individual attacks and deducing statistics about each technique will indicate a lack of security in certain areas. This could, for example, be used for focusing more on techniques that are more frequently used than others. We believe that using the framework for such purposes would benefit the holistic security of mobile communication systems. Similarly, the framework can also be used as a vulnerability impact measurement metrics by adding other dimensions. For instance, techniques in each tactical categories can be arranged according to the complexity or severity of the impact of the techniques. Such newly added dimensions give newer insights when modeling attacks by a specific adversary group and comparing its overall impact with other groups. In these cases, color-coding them further --- \eg, lighter shades for less severe attacks and darker for the most severe ones --- could also be useful. 

As of now, our framework serves as a single point of technical reference for various adversarial techniques used in mobile communication systems. The future directions in which such technical references can flourish include, for example, adding more background and subcategories of the techniques, open-source forensic tools and test suites to detect or defend against the techniques and general best practices and mitigation examples. Tools could also be developed for annotation of the Bhadra matrix and to use such annotation data for deducing further analytics. Similar tools exist for MITRE ATT\&CK framework, which is used regularly by security teams in enterprise networks~\cite{AttckNavigator,AtomicRed}, and we continue to seek inspiration to bring them into the mobile communication sector. To make such developments feasible for the security community, we consider our future contributions to focus on building a repository of attacks modeled with Bhadra and application programming interfaces to access them. Nonetheless, it can only be possible with the help of the community contributions, and we believe that our work initiates such a conversation in the community.

\boldtitle{Limitations.} One of the main limitations of our work is that we deduced the techniques in each tactical categories based on their commonalities across underlying technologies and network subsystems. For instance, ``downgrading'' is a defense evasion technique that is seen in all generations of mobile technology and seen across radio access, core network, and other subsystems. Our methodology is skewed towards such generalization and abstraction of common attacks rather than the attacks that are exclusive to one specific subsystem or technology. Although such generalization is needed to make our framework agnostic to all kinds of attacks, we might have missed the exclusive ones. 

Another limitation of our work is that it is built on the existing attack literature that is publicly available, and it does not include anything that is observed in the wild by mobile operators, which is used exclusively for internal purposes. The former mainly includes attacks that involve UE and radio access network with explicit details from academia, and some of the core network vulnerabilities published as excerpts mostly by the mobile security auditors. There is minimal public information, e.g., for attacks on OSS networks, which we could not fit in our framework. Nonetheless, as and when such information is publicly available, it can be added to our framework. In fact, we encourage entities involved in mobile communication to open up datasets about vulnerabilities observed in the wild for the public good. 




\section{Conclusion}
\label{sec:conclusion}
In this work, we presented a threat modeling framework that is specific to mobile communication systems. Our framework comprises various tactics and techniques that represent an adversary's objectives throughout the life cycle of an attack. We described two use cases of the framework, namely, for modeling individual attacks and for comparing attacks that share some commonalities. We hereby hope that our work unifies prior knowledge and initiates a conversation towards future efforts to secure mobile communication networks.

\bibliographystyle{ACM-Reference-Format}
\bibliography{references}


\end{document}